\documentclass[twocolumn,tighten]{aastex631}

\begin{document}

%\title{Interplay of Planets with Stellar Magnetospheric Accretion: I. Planet Migration and Stochastic Torque Dynamics}
%\title{Young Planets around Young Accreting Stars: I.  Migration and its Effects on Planet Demographics}
\title{Young Planets around Young Accreting Stars: I.  Migration and Inner Stalling Orbits}

\author{Arturo Cevallos Soto}
\affiliation{Department of Physics and Astronomy, University of Nevada, Las Vegas, 4505 S. Maryland Pkwy, Las Vegas, NV, 89154, USA}
\affiliation{Nevada Center for Astrophysics, University of Nevada, Las Vegas, 4505 S. Maryland Pkwy, Las Vegas, NV, 89154, USA}

\author{Zhaohuan Zhu}
\affiliation{Department of Physics and Astronomy, University of Nevada, Las Vegas, 4505 S. Maryland Pkwy, Las Vegas, NV, 89154, USA}
\affiliation{Nevada Center for Astrophysics, University of Nevada, Las Vegas, 4505 S. Maryland Pkwy, Las Vegas, NV, 89154, USA}

\begin{abstract}

%250 word limit
Planet migration within inner protoplanetary disks significantly influences exoplanet architectures.
We investigate various migration mechanisms for young planets close to young stars. To quantify the stochastic migration driven by turbulent disks,
we incorporate planets into existing 3-D MHD disk simulations of magnetospheric accretion. Besides the stochastic torque, we identify periodic torques from slowly evolving disk substructures farther out. We quantify these turbulent torques analytically using a modified Gaussian process. Then, using the disk structure in our simulation, we calculate migration timescales of various processes, including the smooth Type I/II migration, planet-star tidal interaction, magnetic dipole-dipole interaction, unipolar induction, and aerodynamical drag with the magnetosphere.
Since our inner MHD turbulent disk reveals a very low surface density ($\sim 0.01$ g/cm$^{2}$), the resulting disk migration is significantly slower than previously estimated. Earth-mass planets have the migration timescale in the inner MHD turbulent disk exceeding the Hubble time, effectively stalling at the deadzone inner boundary ($R_{\mathrm{DZIB}}$). Only giant planets could migrate inward within the turbulent disk, and may stall at the magnetospheric truncation radius ($R_T$). A simplified planet population synthesis demonstrates that, at the end of the disk phase, all planets around solar-mass stars typically stall at $\lesssim$0.1 au since $R_T\sim R_{\mathrm{DZIB}}$. However, around 2 $M_{\odot}$ stars, higher-mass planets stall significantly closer to the star compared to low-mass planets, due to $R_T\ll R_{\mathrm{DZIB}}$. These results are consistent with recent observations on exoplanet demographics around different types of stars. Finally, turbulence in the low-density disk is unable to break the resonant planets, and thus young planets in resonances may be abundant.

\end{abstract}


\section{Introduction}

When close-in exoplanets form, they likely reside in highly turbulent protoplanetary disks around young T Tauri stars. The inner disk region with temperature higher than $\sim$1000 K is sufficiently ionized to couple with magnetic fields, so that it is subject to the magnetorotational instability to become turbulent \citep[see e.g.,][]{Balbus_1991}.
Meanwhile, this turbulent disk is truncated by the stellar magnetosphere at a few stellar radii (approximately 0.05 --- 0.1 au) \citep[see e.g.,][]{Hartmann_2016}. Material from the inner disk is lifted by magnetic torques, travels along magnetic field lines, and accretes onto the stellar surface at nearly free-fall velocities, impacting at high latitudes \citep[e.g.,][]{Koenigl_1991,Bouvier_2007,armitage2010astrophysics,Bodman_2017}.

% Why we want to study a planet in the magnetosphere region
This turbulent inner disk, undergoing magnetospheric accretion, influences the growth and evolution of close-in young planets. Given the abundance of close-in planets discovered by \textit{Kepler}, the demographics of these exoplanets are well characterized, and can potentially constrain
planet formation and migration mechanisms \citep[e.g.,][]{Howard_2012,Wu_2019}. For example, recent observations have revealed the dependence of the close-in planet population on stellar types, which could be related to different disk properties around these stars \citep[see e.g.,][]{Mendigutia_2024,Sun_2025}. Another example is that the fraction of exoplanets in resonances seems to change with the system's age \citep{Dai_2024}, which put constraints on the disk and exoplanets' dynamical evolution.

However, to bridge planet formation theory with exoplanet observations, we need to understand how planets migrate during and even after the protostellar disk phase. Disk-induced planet migration is due to asymmetric disk features (e.g. spirals) induced by the planet's gravity. The gravitational interaction between these features and the planet provides a torque to the planet, so that the planet migrates towards the inner disk's truncation edge \citep[e.g,][]{PPVI,PPVII}. The most commonly discussed disk-induced torques are Lindblad and corotation torques that lead to Type I and Type II migration \citep[e.g.,][]{Lin1996}. Disk turbulence may provide additional fluctuating torque, leading to stochastic migration \citep[see e.g.,][]{Nelson_2004,Johnson_2006,Adams_Bloch_2009}. Besides planet-disk interactions, other processes can also play significant roles in planet migration, especially for planets in close proximity to their magnetically active host stars. These include star-planet tidal torques (exacerbated by the close proximity) \citep[see e.g.,][]{Hut_1981,Eggleton_1998,Ogilvie_Lin_2004,Ogilvie_Lin_2007,Gallet_2017}, and magnetic torques that arise due to the interaction of unmagnetized stellar wind and the planetary field, dipole-dipole interaction and unipolar induction \citep[see e.g.,][]{Zarka_2007,Laine_2008,Laine_2012,Strugarek_2015,Strugarek_2016}. Furthermore, a conducting or magnetized planet embedded in the changing magnetic field of its host start can induce eddy currents, with the ohmic losses draining the planet's orbital energy over time \citep[e.g,][]{Laine_2008,Kislyakova_2018,Bromley_2022}. The tidal and magnetic interaction could even play important roles after the disk disperses. 

One particularly intriguing possibility regarding planet migration is that migrating planets may stall at some particular radii in the disk. 
Magnetospheric accretion produces a low-density cavity around the star. 
The resulting large density jump at the disk midplane causes the positive corotation torque to dominate over the negative differential Lindblad torque, allowing inward migrating planets to become trapped near the magnetospheric boundary \citep[e.g.,][]{Masset_2006,Romanova_2019}. This planet-trapping mechanism at the magnetospheric boundary is invoked as a way to prevent migrating planets from rapidly inspiraling into the star in planet formation models \citep[e.g.,][]{Morbidelli_2008}. On the other hand, such planet trapping mechanism could operate for any sufficiently steep density jump. Exoplanets could also trap at the deadzone inner edge where the change of accretion efficiency leads to a sharp density jump \citep[e.g.,][]{Chatterjee_2014,Hu2016}.  Understanding these processes and other planet migration mechanisms at the inner disk is thus crucial for the study of close-in exoplanets \citep[e.g.,][]{Lee_2017,Liu_2017}.

% Resonance breaking
Another challenge in understanding exoplanet demographics is explaining why many observed exoplanetary systems do not exhibit mean-motion resonances (MMRs), despite theoretical expectations from disk migration. Convergent migration of planets within the protoplanetary disk tends to bring planetary pairs into orbital configurations where their period ratios are simple fractions of consecutive integers, due to the trapping in MMRs \citep[e.g,][]{Goldreich_1965,Allan_1969,Sinclair_1970,Sinclair_1972}. Nevertheless, the overall distribution of exoplanetary orbital periods shows little preference for such commensurabilities \citep[e.g.,][]{Fabrycky_2014}. It is possible that turbulence within the disk stochastically disrupts previously established resonances between planet pairs, scattering them and leading to orbital architectures that are non-resonant. This can provide a possible explanation for the observed diversity in exoplanetary system configurations \citep[e.g.,][]{Rein_2012,Batygin_2017}.

% 3-D simulations
% Previous results from ZSC24 - in general, migration and detection
The migration of young planets and the break of the resonance chains are largely determined by the properties of the young protoplanetary disks. Direct numerical simulations have been carried out to study the magnetospheric accretion of the inner disk \citep[see review by][]{Romanova_2015}, which reveals complex disk structures. For example, the recent high-resolution 3-D ideal MHD simulations by \cite{zhu2023global} (hereafter referred to as ZSC24) reveal a highly turbulent inner protoplanetary disk. For slow rotators, filamentary flows emerge at the magnetospheric truncation radius ($R_T$), some climbing the magnetosphere and some penetrating through the midplane. Highly magnetized large-scale bubbles, similar to those in magnetically arrested disks (MAD) around black holes \citep[e.g.,][]{Parfrey_Tchekhovskoy_2023}, form recurrently at the magnetospheric truncation radius due to magnetic reconnection and "interchange instability". On the other hand, stellar rotation suppresses this instability and filamentary flow at the midplane, leaving a clean magnetosphere cavity \citep{Zhu_2025}.

In this paper, we first use 3-D numerical simulations to study planet migration in a disk undergoing magnetospheric accretion, then discuss various migration mechanisms using our realistic inner disk structure, and finally examine how these planet migration processes affect the demographics of exoplanets. 

In \S\ref{sec:theo}, we outline unified Type I/II migration and introduce the concept of stochastic torque-driven diffusion, defining key parameters \(\tau_c\) and \(C_D\) and explaining how they can be constrained by simulations. In \S\ref{sec:numerical} the numerical model is presented, with special emphasis on the addition of orbiting planets to it. The results obtained from simulations are presented in \S\ref{sec:results} with a focus on the interpretation and modeling of the planetary torque, along with measurement of parameters \(\tau_c\) and \(C_D\), and a planet's influence on the accretion rate signature of their host star. In \S\ref{sec:disc} we compare and discuss different planetary migration timescales, assess their impact on close-in planet demographics, examine the stability of resonant planet pairs, and assess the representativeness of our simulations. We summarize our work and share our conclusions in \S\ref{sec:conclusion}. Supplementary results are presented in Appendix \S\ref{sec:appendix}.

\section{Theoretical Framework}\label{sec:theo}

\subsection{Unified Type I and Type II Migration Timescale}

Planet formation models suggest that many exoplanets do not form \textit{in situ}, but rather originate at larger radii in protoplanetary disks and migrate inward due to gravitational interactions with the disk \citep[e.g.,][]{Lin_1986, Kley_2012}. When a low-mass planet (typically less than a few Earth masses) is embedded in a gaseous disk, it undergoes so-called Type I migration, driven by the differential Lindblad and corotation torques exerted by the surrounding disk material \citep[e.g.,][]{Tanaka_2002,Tanaka_2004}. As the planet mass grows, it can carve a partial or full gap in the disk, transitioning into the Type II regime, where migration generally slows and is coupled to the viscous evolution of the disk \citep{Lin_1986}.

ZSC24 combined Type I and Type II migration timescale into one equation. Specifically, for a planet of mass ratio \(q = M_p/M_{\star}\) at orbital radius \(r_p\) in a disk with surface density \(\Sigma\), aspect ratio \(h = H/r\), and dimensionless viscosity parameter \(\alpha\), the migration timescale (see Equation 41 in ZSC24, \citealt{Dempsey2020}) is given by
\begin{equation}
    t_{\mathrm{mig}} = \Omega^{-1} h^2 q^{-1} \left( \frac{\Sigma r_p^2}{M_{\star}} \right)^{-1} (1+hK),
\label{eq:disk-driven}
\end{equation}
where $K \equiv q^2/(\alpha h^5)$, $\Sigma$ is the background disk surface density without the planet, and all quantities are evaluated at the planet's location.

In this formulation, when \(K \lesssim 1/h\), the migration rate reduces to the standard Type I regime. Massive planets enhance \(K\) which puts them in the Type II regime and slows their migration compared to Type I. When a gap is opened, a higher \(\alpha\) value enhances the disk's radial accretion velocity, accelerating Type II migration \citep[e.g.,][]{Crida_2006}.

\subsection{The Turbulent Diffusion Timescale}

In addition to Type I or II migration, magnetorotational instability (MRI) turbulence introduces stochastic torques on embedded planets \citep{Nelson_2004, Baruteau_2010}. These random fluctuations in the disk density and velocity field impart a random-walk on the planet's semi-major axis. Earlier approaches \citep[e.g.,][]{Laughlin_2004, Johnson_2006} assume the correlation time \(\tau_c\), the typical timescale of fluctuations, to be approximately half an orbital period. With a circular orbit assumption, the correlation time is thus:
\begin{equation}
    \tau_c \approx \frac{\pi}{\Omega} = \frac{\pi (J/M_p)^3}{(G M_{\star})^2},\label{eq:taucadam}
\end{equation}
where \(J\) is the orbital angular momentum \(J=M_p (G M_{\star} r_p)^{1/2}\) of the planet.

For a given torque \(\Gamma\), whose fluctuating component is \( \delta \Gamma \), the variance of the torque's fluctuations is taken to be
\begin{equation}
    \overline{[\delta \Gamma (t,J)]^2} \approx (C_D \Gamma_0)^2 = (C_D 2 \pi G \Sigma M_p r_p)^2,
\label{eq:fluctuations}
\end{equation}
where $C_D$ is a dimensionless coefficient depending on the strength of the turbulence; $\Gamma_0 = 2 \pi G \Sigma M_p r_p$ is the natural torque scaling calculated using the force that a planet would feel if suspended above the disk.

Combining the torque's variance and the correlation time $\tau_c$, we arrive to a diffusion coefficient, which describes the efficiency and rate at which angular momentum is randomly exchanged within the disk due to turbulent motions:
\begin{equation}
    D(J) = \tau_c \overline{[\delta \Gamma (t,J)]^2}.
\end{equation}

The corresponding diffusion timescale \(t_{\mathrm{diff}}\) measures the characteristic time over which stochastic processes significantly alter the planet’s orbital angular momentum. With \( t_{\mathrm{diff}} = J^2/D \), we have
\begin{equation}
    t_{\mathrm{diff}} =  \frac{M_*}{4\pi^2  \tau_c C_D^2 G\Sigma^2 r_p}\,.
\label{eq:diffusion0}
\end{equation}
After plugging in Equation \ref{eq:taucadam}, we have
\begin{equation}
    t_{\mathrm{diff}} =  \frac{1}{4 \pi^3 C_D^2 \Sigma^2} \left( \frac{M_{\star}^3}{G r_p^5} \right)^{1/2}.
\label{eq:diffusion}
\end{equation}

A notable property of the diffusion timescale is that, when compared to the Type I/II timescale (Equation \ref{eq:disk-driven}) and tidal migration timescale, the turbulent diffusion timescale does not depend on the planet's mass. However, the values of both \(\tau_c\) and \(C_D\) remain poorly constrained by theory alone. Different works have adopted simplified assumptions or order-of-magnitude estimates, with both \cite{Johnson_2006} and \cite{Adams_2009} taking $\tau_c=\pi/\Omega$ and assuming $C_D$ to be independent of radius and on the order of \(C_D \sim 0.05\). One of our goals in this work is, through analyzing the torque time series in the simulation, we could measure its autocorrelation function and thereby determine \(\tau_c\), and by comparing the torque amplitude to \(\Gamma_0\), we determine \(C_D\) at different radial positions.

To proceed, we assume an exponential decay for the torque autocorrelation function, consistent with some numerical studies of MRI turbulence \citep{Fromang_2006, Oishi_2007}. Following \citet{Rein_2009}, if \(F_{\phi}(t)\) is the azimuthal force per unit mass acting on the planet, we write:
\begin{equation}
    \langle F_{\phi}(t) F_{\phi}(t') \rangle = \langle F_{\phi}^2 \rangle \, g(| t - t' |),\label{eq:auto}
\end{equation}
where \(\langle F_{\phi}^2 \rangle \) is root-mean-square (RMS) value, and \(g(\Delta t)\) is the autocorrelation function with

\begin{equation}
    g(\Delta t) = \exp\left( -\frac{\Delta t}{\tau_c} \right). \label{eq:exp_fit}
\end{equation}
and
\begin{equation}
    \int_0^{+\infty} g(\Delta t) \, \mathrm{d}\Delta t = \tau_c,
\end{equation}

Once the RMS value and \(\tau_c\) have been determined, the observed stochastic nature of the torque can be modeled by using a simple one-dimensional Markov chain \citep{Kasdin_1995, Rein_2009} (Section \ref{sec:resultsmark}).

\section{Numerical Method}\label{sec:numerical}

The simulation setup for the disk, the star and the initial conditions closely follows the methodology outlined in ZSC24, where magnetohydrodynamic (MHD) equations are solved in the ideal MHD limit using Athena++. Static mesh-refinement has been adopted. We summarize the key setup briefly. For further details, see Section 3 in ZSC24. 

\subsection{Disk Setup}

Close to the star, stellar magnetic fields are so strong that the magnetic pressure becomes larger than the accretion ram pressure inside the Alfvén radius $R_A$, defined as
\begin{equation}
    R_A = R_{\star} \left( \frac{B_{\star}^4 R_{\star}^5}{2 G M_{\star} \dot{M}^2}  \right)^{1/7} \:{\rm in \: C.G.S.,}
\label{eq:truncation}
\end{equation}
where \(B_{\star}\) is the stellar magnetic field at the equator, \(R_{\star}\) is the stellar radius, and \(\dot{M}\) is the accretion rate. In ZSC24, $R_A \sim R_T$, where \(R_T\) is the magnetospheric truncation radius, which is defined as the radius where the disk's azimuthal velocity drops to half the Keplerian velocity $v_K$. It also corresponds to where the disk density changes sharply. $R_0$ is the code length unit, which is quite close to $R_T$ at the end of the simulation. The stellar radius, denoted as \(R_{\star}\), is chosen as $0.1~R_0$, so that the magnetospheric truncation radius is roughly 10 times larger than it.

The initial disk has a midplane density profile as
\begin{equation}
    \rho_d(r,z=0) = \rho_0 \left( \frac{r}{R_0} \right)^p,
\label{eq:den1}
\end{equation}
with $\rho_0 \equiv \rho_d(r=R_0,z=0)=1$. The time unit $1/\Omega(r=R_0)=T_0/2\pi$, in which $T_0$ is the orbital period at $R_0$.

The temperature is set to be constant at each cylindrical radius:
\begin{equation}
    c_s^2(r,z) = c_s^2(r=R_0,z=0)\left( \frac{r}{R_0} \right)^q,
\end{equation}
where $c_s=\sqrt{p/\rho}$ is the isothermal sound speed. By choosing $p=-2.25$ and $q=-1/2$ we get a disk surface density $\Sigma \propto r^{-1}$. For the scale height $H=c_s/\Omega_K$, we set $(H/r)_{r=R_0}=0.1$.

We adopt a locally isothermal equation of state using
a fast cooling. More details for this treatment can be found in \cite{Zhu_2015}.

\subsection{Star Setup}

For the central object, we introduce a stellar structure with a gravitational pull outside the radius \(R_{\star}\), and a fixed density, velocity structure and pressure when at \(r<R_{\star}\). The star is set to be non-rotating. The magnetic field is set to be a dipole. In Athena++ the vacuum permeability constant is set to 1 so that the magnetic pressure is simply $B^2/2$ in code units. Values are chosen so that the initial plasma $\beta = 2P/B^2$ at $r=R_0$ is 10. A density floor that varies with position is introduced in the highly magnetized magnetosphere to avoid small timesteps.  For further details see Section 3.2 of ZSC24.

Although our simulation adopts a non-rotating star, we will also discuss planetary migration around rotating stars in Section \ref{sec:disc}. For a rotating star, we can define a corotation radius in the disk, \(R_C\), where the disk's Keplerian frequency matches the star's spin rate. For stars in the spin equilibrium state, simulations by \cite{Zhu_2025} find \(R_T = 0.79~R_C\)  which places $R_C$ a bit outside the magnetospheric cavity.

For the rest of the paper, we drop the code unit (e.g., \(\rho_0\), \(R_0\), \(T_0\)) in expressions for ease of reading.

\subsection{Planet Setup}

\begin{figure*}[t!]
\centering
\includegraphics[trim=0mm 0mm 0mm 0mm, clip, width=7.in]{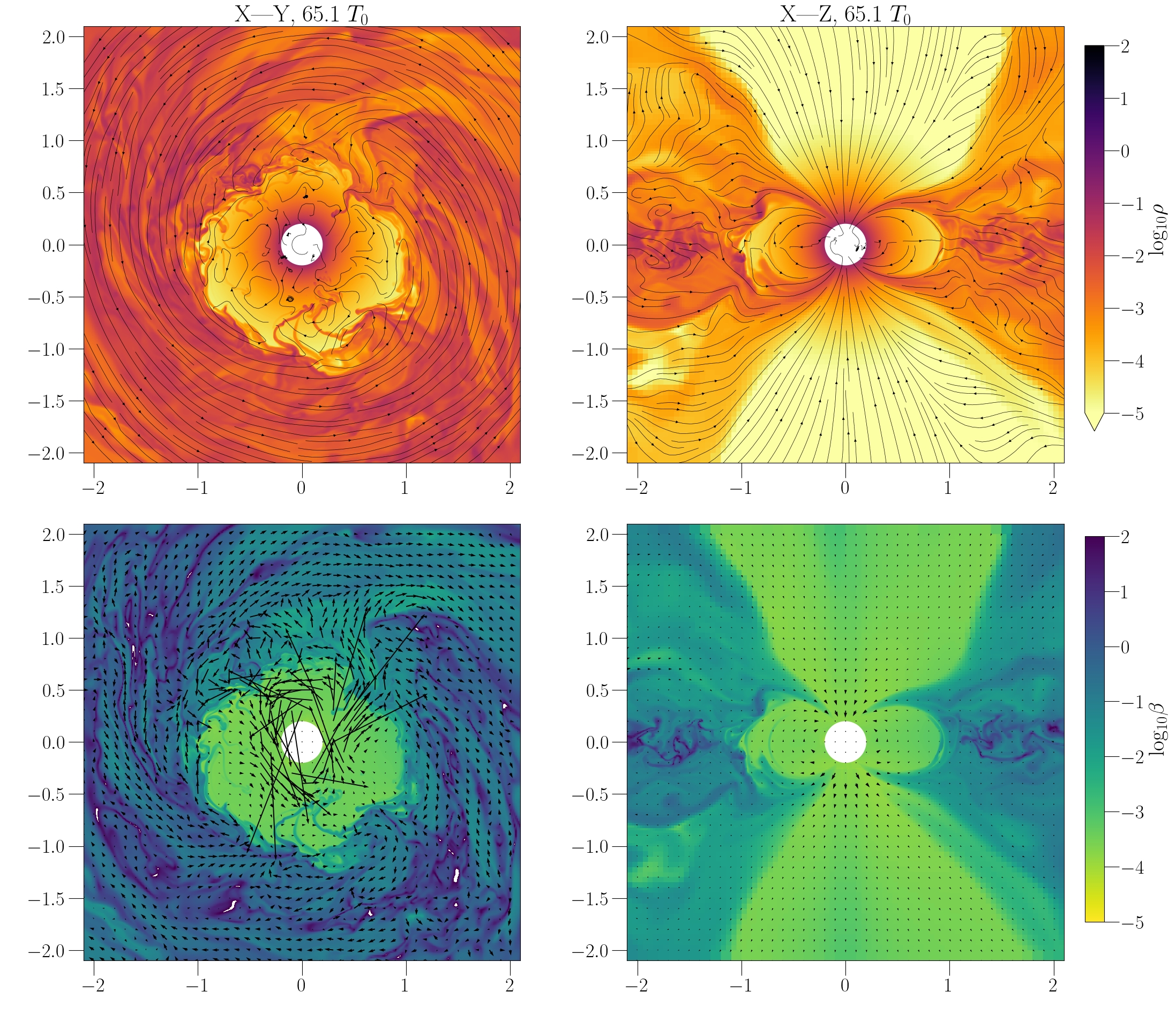}
\caption{Midplane (left) and poloidal slices (right) for density (upper) and plasma $\beta$ (bottom) at the beginning of the simulation. Velocity streamlines and magnetic vectors are overplotted on the upper $\rho$ and bottom $\beta$ panels, respectively.\label{fig:polmidslice}}
\end{figure*}

% Introductory context
Planets are introduced after the disk has reached a quasi-steady accretion stage. We have chosen $t = 65.0$ or the 65th orbit at $r=R_0$ in ZSC24 as our starting time. The state of the disk just before the planet is introduced is shown in Figure \ref{fig:polmidslice}. As in ZSC24's simulations, within the magnetosphere region of \(r<1\), disk material becomes filamentary, with gas being clumped together into high density accretion columns or "fingers" due to interchange instability. While some columns follow the magnetic field lines, being lifted up from the midplane and falling onto the star at about \(30^{\circ}\) from the poles, some could penetrate deep into the magnetosphere at the midplane. Thus, while the region inside the magnetospheric boundary has very low density and is fully magnetically dominated \(\beta<<1\), it is not completely empty. When the star rotates sufficiently fast (even faster than the equilibrium spin state), the interchange instability will be suppressed and the cavity will be clean \citep{Zhu_2025}. 

% Representation and orbit details
Planets are added into the disk as orbiting gravitational potentials. The mass ratio between the planet and the central star is $q$. 
For the planet's gravitational potential, we apply a smoothing radius \(R_s = 0.012\). This smoothing length approximately corresponds to Jupiter's radius considering that the stellar surface is at $R_*$=0.1. This value is small enough to accurately capture the dynamical interactions between the planet and the nearby disk material.

% Case definitions
For our simulations, we consider two distinct cases:
\begin{itemize}
    \item Case A: Massive planet with \(q=0.01\), equivalent to \(10~M_J\), located at \(r_p=1\).
    \item Case B: Three smaller planets with \(q = 10^{-4}\), each roughly equivalent to \(30~M_{\oplus}\), located at \(r_p=0.45\), \(0.9\) and \(1.8\), respectively.
\end{itemize}

In all these cases, the planets follow fixed circular orbits at the local Keplerian velocity \(v_K\).

In case A we have a very massive planet in the system which may be able to moderate the accretion from the disk towards the star; a process which could potentially be detectable via observations of variable near-UV excess from accretion shocks or variable $H_{\alpha}$ line as the disk material undergoes magnetospheric accretion \citep[e.g.,][]{Hartmann_2016}. Meanwhile, the low-mass planets in Case B will not affect the disk structure, allowing us to study the interaction between the disk and multiple planets at different radii simultaneously in one simulation. 
We thus put three planets inside the magnetosphere, close to the truncation radius, and embedded in the disk, respectively to study the planetary torque and migration behavior for the same type of planet at different starting conditions. 

% Torque calculation
To measure the planetary torque, we exclude the mass within the planet's Hill sphere as it is gravitationally bound to the planet and can be considered as part of it \citep[e.g.,][]{Ben_tez_Llambay_2015}. Without sufficient resolution, the circumplanetary region is not well resolved. Due to its proximity, any structure in this region leads to considerable fluctuations when calculating the torque. For the same reason we remove the innermost cells around the star, excising the volume within 2.5~\(R_{\odot}\). During each time snapshot, we measure the contributions of each cell toward the net force vector felt by the planet. Then, we transform the net force's Cartesian components to the components in spherical coordinates and get the azimuthal component \(F_{\phi}\): the planetary torque which is a main driver for planetary migration \citep[][]{Kley_2012,Lesur_2021}.
%\todo{Please give specific radius where it is excluded}
%We also exclude the volume occupied by the star itself up to 2.5~\(R_{\odot}\) for the same reason.

\section{Results}\label{sec:results}

%%% Visualization figure
\begin{figure*}[t!]
\centering
\includegraphics[trim=0mm 0mm 0mm 0mm, clip, width=7.in]{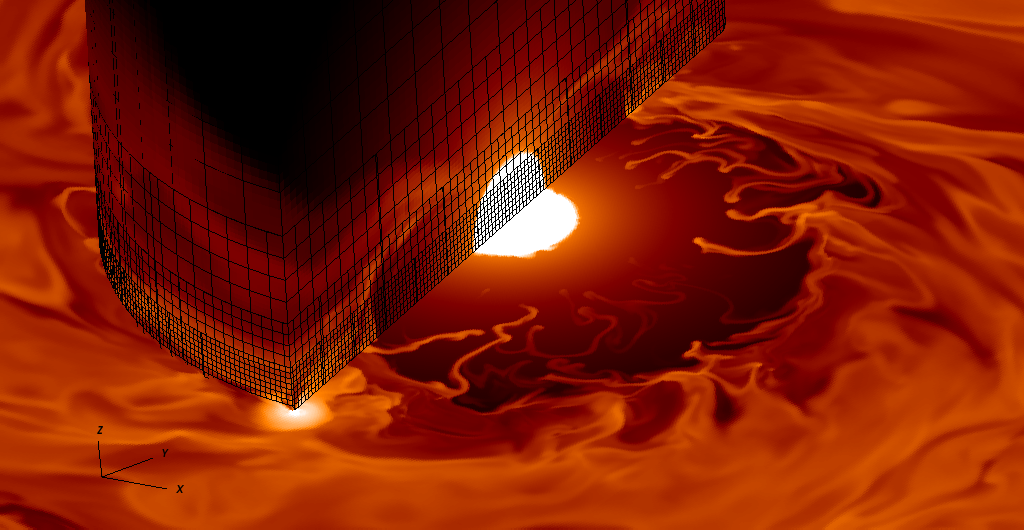}
\caption{Color-contour visualization of density at the disk midplane (the X-Y plane), combined with a vertical slice (the x-z plane), and a cylindrical segment ($\phi$ coordinate at $r=1$) at time $t=88.35$. These three surfaces intersect at the current location of the massive $q=0.01$ planet. The adopted mesh structure is also shown as an overlay on the half-cylinder segment.}
\label{fig:visit}
\end{figure*}

Limited by the computational resources, we run the simulations for tens of orbits at the truncation radius. Each simulation produces 3-D snapshots that contain the density, velocity, and magnetic fields of each cell. For the fiducial one-planet system (case A), we are able to get close to 30 orbits at \(r=1\), with 99 snapshots each separated by 0.1 orbit, and a further 368 snapshots separated by 0.05 orbit. For the triple planet system (case B), we obtain five orbits total, with 101 snapshots separated by 0.05 orbit. For every case, aggregate quantities, like the net force exerted on each planet by the entire disk, and the planets' respective positions, are saved in a history-file at 10 times finer time resolution than the 3-D snapshots.

Figure \ref{fig:visit} shows color contours of density and the grid structure near the star for a single snapshot of case A. The vertical slices in the figure are deliberately placed to intersect the cylinder of radius $r=1.0$ at the planet's current location, highlighting the planet's impact on the surrounding gas. Upon introduction, the massive planet quickly begins attracting nearby material. The fate of this accreted gas, whether it remains in a spherical envelope or forms a circumplanetary disk, will be the subject of a future work.

% What we will talk about
In the following subsections we will present the planetary torque time series, and develop a quasi-stochastic mathematical model that satisfactorily reproduces the time series. This allows us to constrain the coefficients of turbulent diffusive migration.

\subsection{Planetary Torque}\label{sub:planet_torque}

%%% Torque figure %%%
\begin{figure*}[t!]
\centering
\includegraphics[trim=0mm 0mm 0mm 0mm, clip, width=7.in]{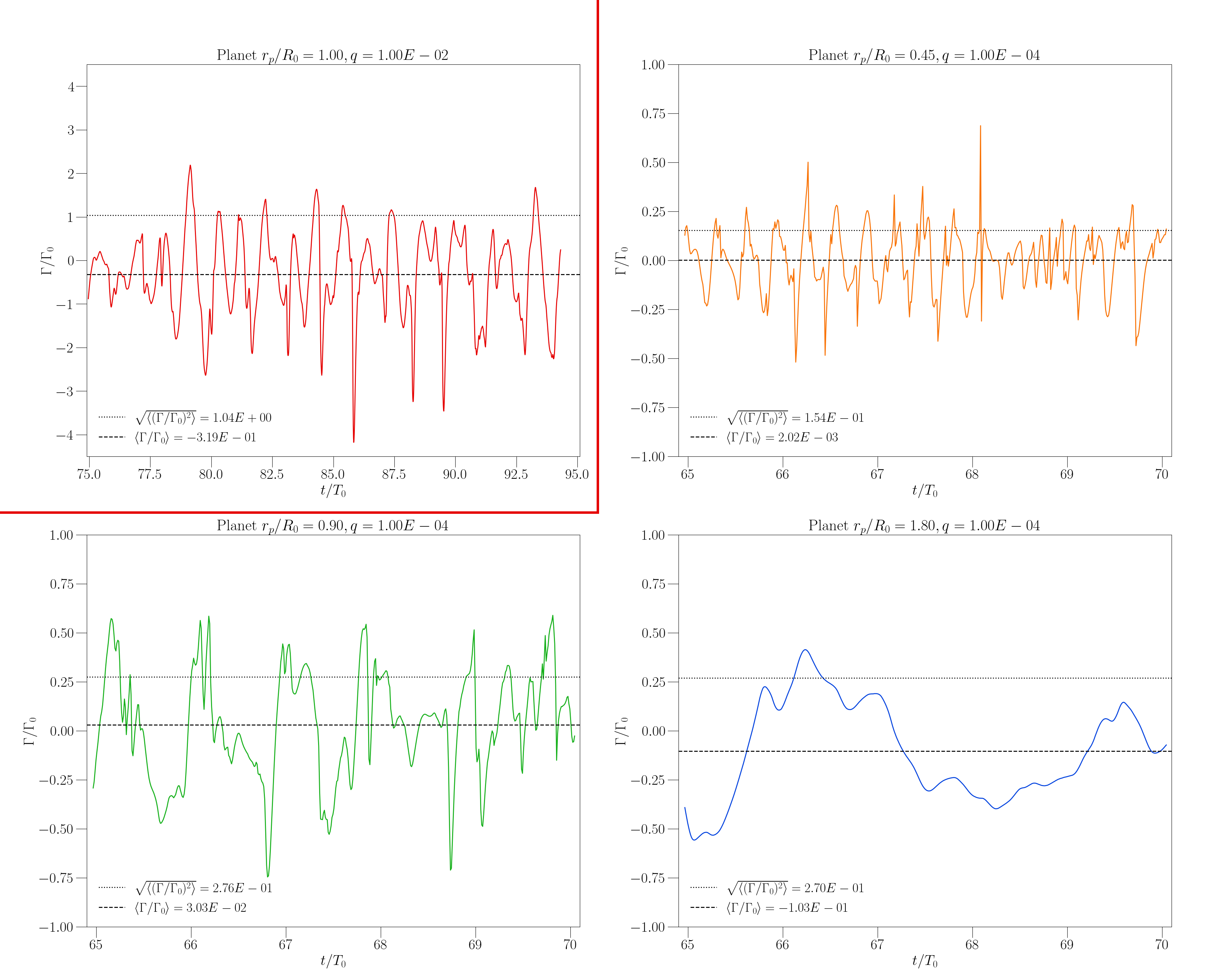}
\caption{Total normalized torque on each planet (solid line), its root-mean-square value (RMS, dotted line), and avera (dashed line) with respect to time. \textbf{Top left:} single $q=0.01$ planet system. \textbf{Rest:} triple $q=10^{-4}$ planets system.}\label{fig:torque_t4_t2}
\end{figure*}

% What is torque and that Gamma and F_phi are the same thing.
The planetary torque represents the gravitational torque exerted on a planet by the surrounding protoplanetary disk. It dictates the angular momentum exchange between planets and disks, which leads to the changes in the planet's orbital radius over time \citep{Papaloizou_2006}.\footnote{We have neglected the magnetic torque which will be studied in future.}
During our simulations, we measure the net torque acting on each planet with time and label it as \( \Gamma \). Since our planets are on circular orbits with no inclination or eccentricity, only the azimuthal (tangential) component of the gravitational force, $F_{\phi}$, would contribute to changes in the planet's angular momentum. The resulting torque is then normalized by the natural scale of the turbulent torque \(\Gamma_0\) as in Equation \ref{eq:fluctuations} \citep[e.g.,][]{Johnson_2006,Rein_2009}. \(\Gamma_0\) uses the initial azimuthally averaged surface density at each location.
%\todo{What is the time average mean here? Did you average over the whole simulation time?} - DONE. I ended up using the initial value because the large planet does affect the disk structure over time.

Figure \ref{fig:torque_t4_t2} shows the normalized torques \(\Gamma/\Gamma_0\), felt by the single large planet of case A and the three planets of case B, along with their corresponding root-mean-square (RMS) value, which quantifies the typical magnitude of fluctuations. For the $q=0.01$ planet, the RMS value is close to unity - about four times larger than the values measured for the low-mass \(q=10^{-4}\) planets. If the turbulent disk is not affected by the planet, the normalized torque should be independent from the planet mass. Thus, the larger RMS value for the massive planet indicates that the massive planet has some feedback to the background turbulent disks. One possible feedback is gap opening by the massive planet \citep[e.g.,][]{Oishi_2007}. However, with the strong turbulence at the inner disk, a gap has not been induced by the planet. We suspect that the feedback arises from the magnetic interaction between the circumplanetary region and the disk, which deserves further study in future. Nevertheless, the two orders of magnitude change in planet mass only leads to a factor of 4 difference in RMS.

In addition to the stochastic torque, the planet should also feel a net Type I/II migration torque due to its interaction with the launched spirals \citep[see][]{Zhu_2015}.
Thus, we average the torque profiles to derive the mean torques felt by the planets (shown in Figure \ref{fig:torque_t4_t2}). If we normalize the mean torque with the typical Type I torque \(\Sigma_0 r_p^4 \Omega_p^2 q^2 (c_0 /v_p )^{-2}\) where \(c_0\) and \(v_p\) are the sound speed and the planet's orbital speed, the massive planet ($q=0.01$) at $r_p$=1 registers an average normalized torque of $-0.92$, indicating a gentle inward pull. For the $q=10^{-4}$ planets, the innermost one yields $+0.57$, the one at $r_p = 0.9$ shows a very strong outward torque of $+17.12$, and the outermost planet registers $-116.3$. Although this implies that the zero torque radius is between 0.9 and 1 which is consistent with trapping by the truncation radius, these averaged torques are strongly affected by the statistical error, not reflecting the true mean Type I torque. The ratio between the stochastic torque ($\sim 2\pi G\Sigma M_p r_p$) and the Type I torque ($\sim \Sigma r_{p}^4\Omega_p^2 q^2 h^{-2}$) is on the order of $2\pi h^2/q$.  With \(h=0.1\), the Type I torque is smaller than the stochastic torque by a factor of $\sim$ 6 for a \(q=0.01\) planet, and a factor of $\sim$ 600 for a $q=10^{-4}$ planet. To derive the small mean torque from the larger stochastic torque, we thus need 36 and 3.6$\times 10^5$ independent torque measurements. Since the stochastic torque is correlated within the local orbital time (discussed below), the required simulation duration, especially for the low mass planet or planets further from the star, is much longer than our simulation time. 

% Have to mention the autocorrelation function exponential model here
To analyze the temporal behavior of the stochastic torque, we calculate the autocorrelation function (Equation \ref{eq:auto}) for the torque time series. This analysis is carried out by comparing the torque dataset with a lagged version of itself. For a given lag, the data is shifted, and the pointwise product of the original and shifted datasets is computed. The time average of these products is then normalized by the variance of the dataset, ensuring the autocorrelation values range between -1 and 1. This process is repeated for different lag values to determine the autocorrelation across various timescales, called the autocorrelation function (ACRF).

%%% Autocorrelation
\begin{figure*}[t!]
\centering
\includegraphics[trim=0mm 0mm 0mm 0mm, clip, width=7.in]{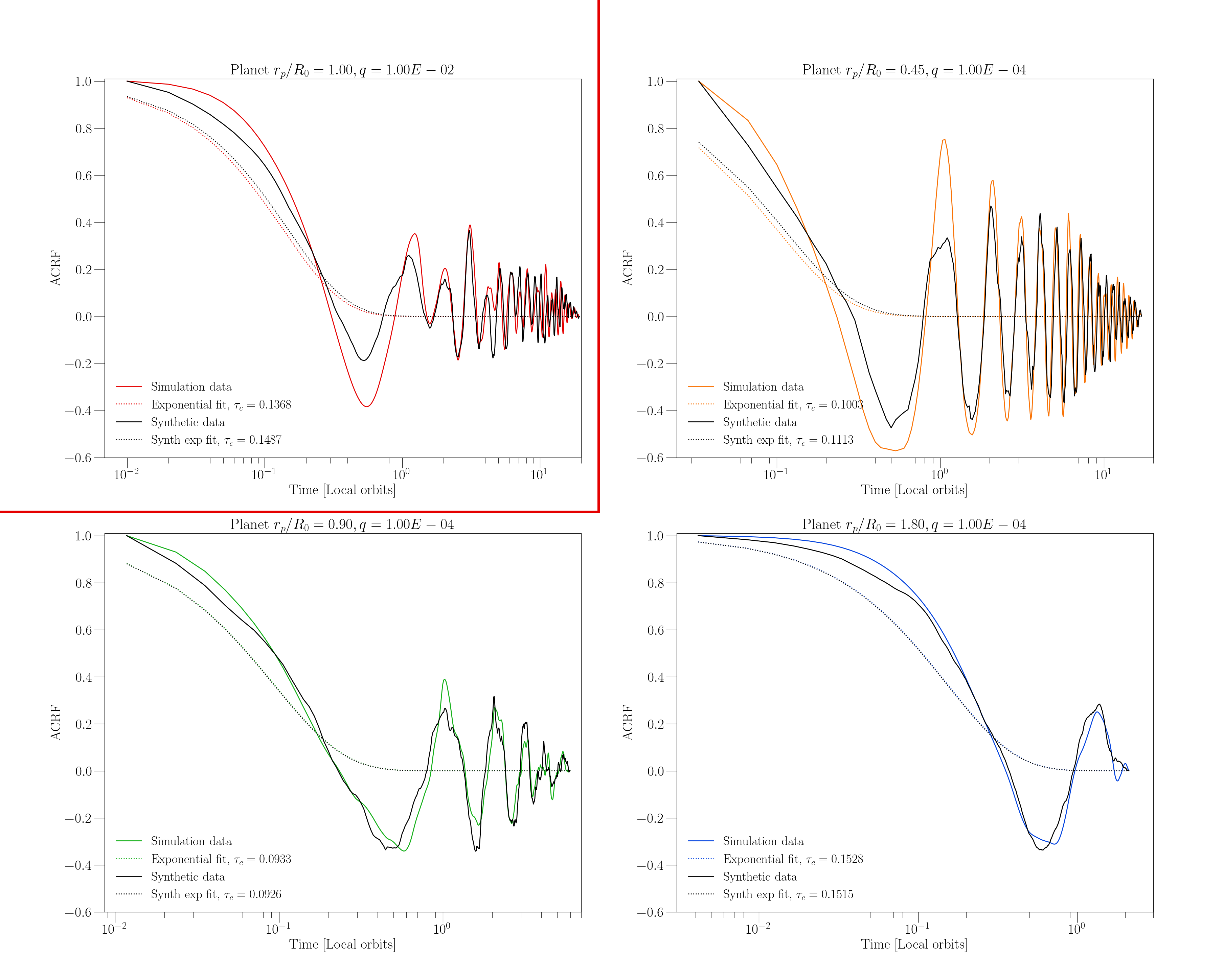}
\caption{The autocorrelation functions and their exponential fits for each of the measured planetary torques (color lines). The best-fitting synthetic torques' autocorrelation functions and their respective exponential fits are also plotted (black lines). Time is measured in local orbits.}\label{fig:autocorrelation}
\end{figure*}

The ACRFs of each planet, along with their corresponding exponential fits, as defined in Equation \ref{eq:exp_fit}, are presented in Figure \ref{fig:autocorrelation}. Each graph also includes an  ACRF from our synthetic torque and the corresponding exponential fit, which will be addressed in the following subsection.

In each case, the correlation time, \(\tau_c\) (written on each graph and also in Table \ref{tab:coef}), is found to be on the order of \(\approx 0.1 \mathrm{ - } 0.2\) times the local orbital period. This aligns with the rapid torque fluctuations experienced by planets embedded in a highly turbulent flow, driven by MRI (see Figure \ref{fig:torque_t4_t2}), and thus \(\tau_c\) also represents the eddy turnover time of MRI turbulence. 

One new feature we notice in the autocorrelation function is that it shows high positive peaks that appear approximately every local orbit, suggesting that the torque felt by each planet has a periodic component or "tug" that corresponds to the orbital period. Such a correlation timescale is absent in local shearing box MRI simulations \citep[e.g.,][]{Oishi_2007}. This implies that some relatively long-lived global disk structures persist over a few orbits, imprinting a characteristic periodicity on the torque felt by the planet each orbit. 

%%% Segmented disk torque and autocorr
\begin{figure*}[t!]
\centering
\includegraphics[trim=0mm 0mm 0mm 0mm, clip, width=7.in]{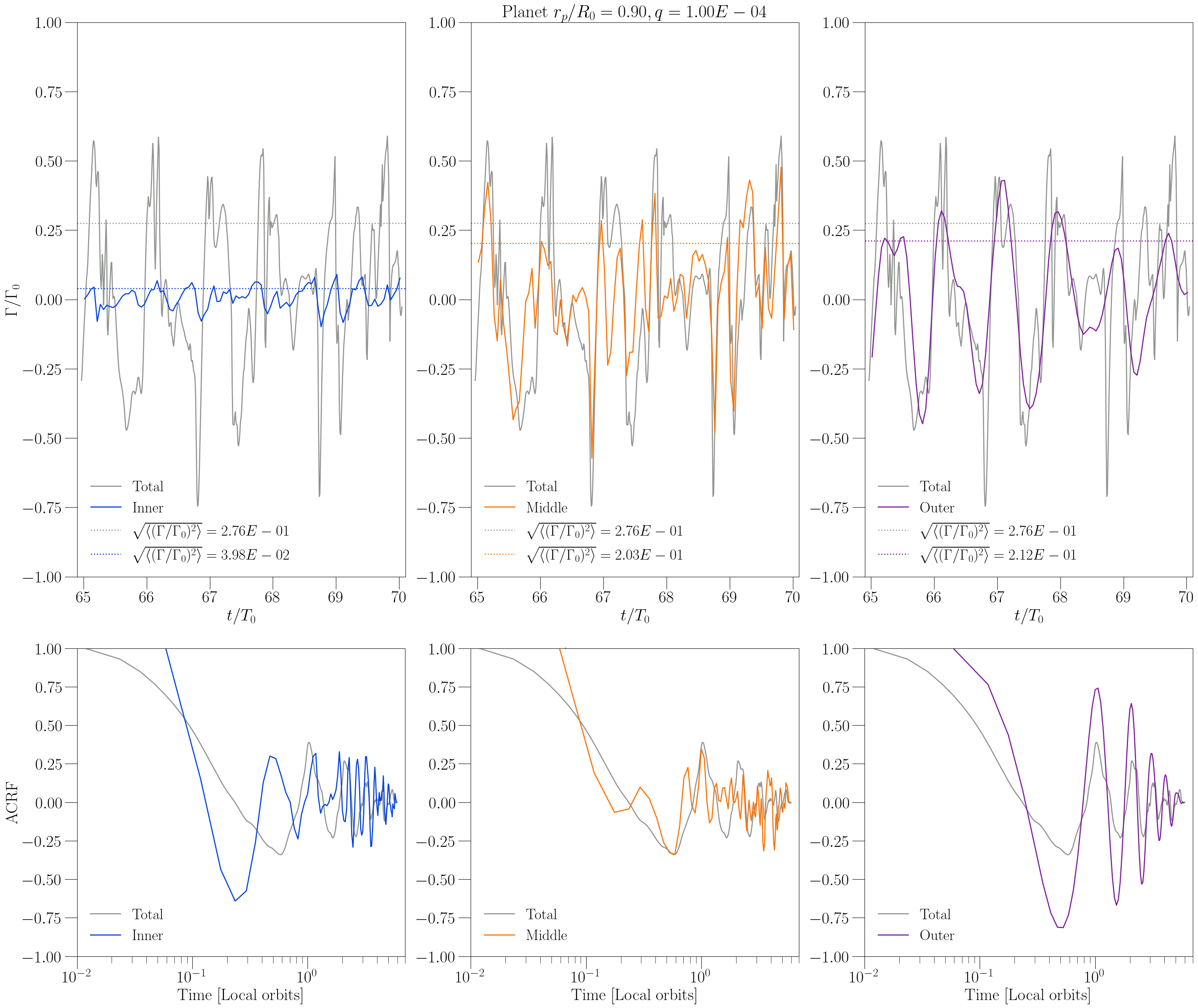}
\caption{\textbf{Top:} Normalized planetary torque felt by the second planet in case B (\(r = 0.9\), \(q=10^{-4}\)), arising from different annuli of the accretion disk versus the total. \textit{Inner}, \textit{Middle}, and \textit{Outer} sectors correspond to annuli at \(r \leq 0.8\), \(0.8 < r \leq 1.5\), and \( r > 1.5\), respectively. \textbf{Bottom:} Autocorrelation functions of the sectoral torque.}\label{fig:segment-1}
\end{figure*}

To further explore where most torque comes from, we present the planetary torque and its autocorrelation like those shown on Figures \ref{fig:torque_t4_t2} and \ref{fig:autocorrelation} but for individual segments of the disk. The disk has been divided into three parts: an inner, a middle, and an outer part; they correspond to regions at \(r \leq 0.8\), \(0.8 < r \leq 1.5\), and \( r > 1.5\), respectively. We take the measurements of the torque arising from each part and compare it to the full-disk measurement, while also calculating the ACRFs. In Figure \ref{fig:segment-1} we show these torque contributions as felt by the second planet in case B (\(r_p = 0.9\), \(q=10^{-4}\)), and their respective ACRFs. This analysis demonstrates that, for this planet located close to the truncation boundary, most of the torque is from the middle and outer regions of the disk, while only a tiny portion from the inner, mass-depleted, magnetosphere cavity, despite the presence of the dense accretion columns. However, regardless of the magnitude of the torque from each part, the periodic behavior previously observed in the full-disk analysis clearly arises from the outer disk, with its ACRF's peaks being of the highest magnitude, and appearing clearly on each orbit. The same analysis for case B's other planets, \(r_p = 0.45\) and \(r_p = 1.8\), can be found in Appendix \S\ref{sec:appendix}.

\begin{figure*}[t!]
\centering
\includegraphics[trim=0mm 0mm 0mm 0mm, clip, width=7.in]{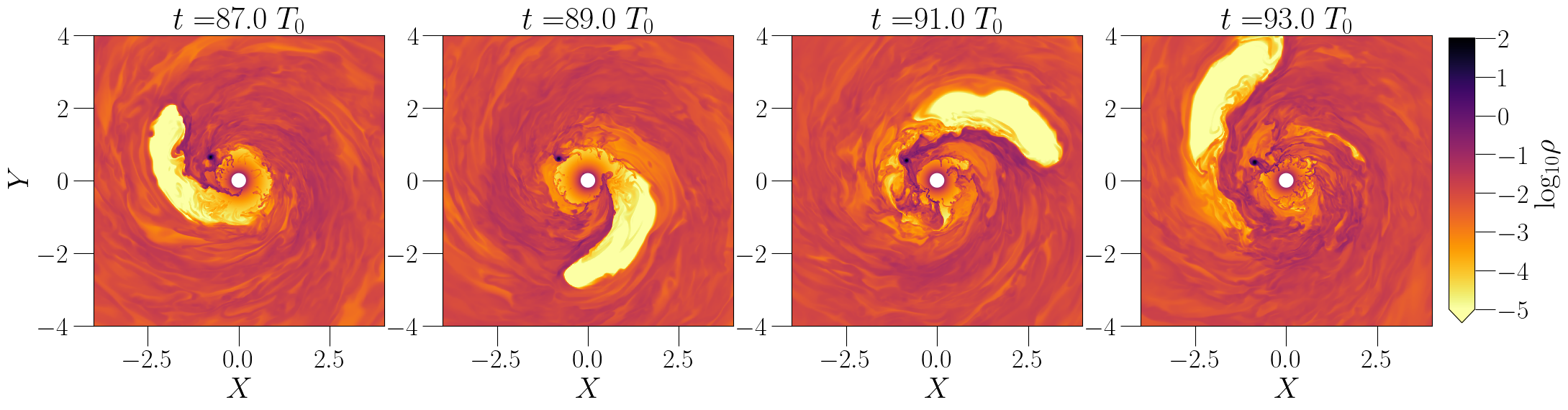}
\caption{Density map of the midplane at different times for the case A simulation. The \(q=0.01\) planet can be discerned as a purple dot at \(r=1\). The growth and dynamics of the magnetically dominated bubble are clearly illustrated.}\label{fig:bubble_t4}
\end{figure*}

%Statements about continuos component of the torque
As we will see in the next subsection, the periodic component of the torque is small in magnitude compared to the stochastic component. However, its periodic character is reminiscent of the time-varying perturbations experienced by circumbinary planets \citep[e.g.,][]{Blundell_2011} or stars in barred galaxies \citep[e.g.,][]{Contopoulos_1980}, where the non-axisymmetric gravitational potential induces periodic modulations in orbital parameters such as eccentricity and precession. Based on the midplane density contours in Figure \ref{fig:polmidslice}, the periodic torque is likely from the eccentric cavity or asymmetric disk features due to magnetic bubbles \citep{zhu2023global}. An example of one such bubble’s formation and evolution is shown in Figure \ref{fig:bubble_t4}.

%%% Bubble figure %%%

\subsection{Synthetic Torque}
\label{sec:resultsmark}
%%% Synthetic torque
\begin{figure*}[t!]
\centering
\includegraphics[trim=0mm 0mm 0mm 0mm, clip, width=7.in]{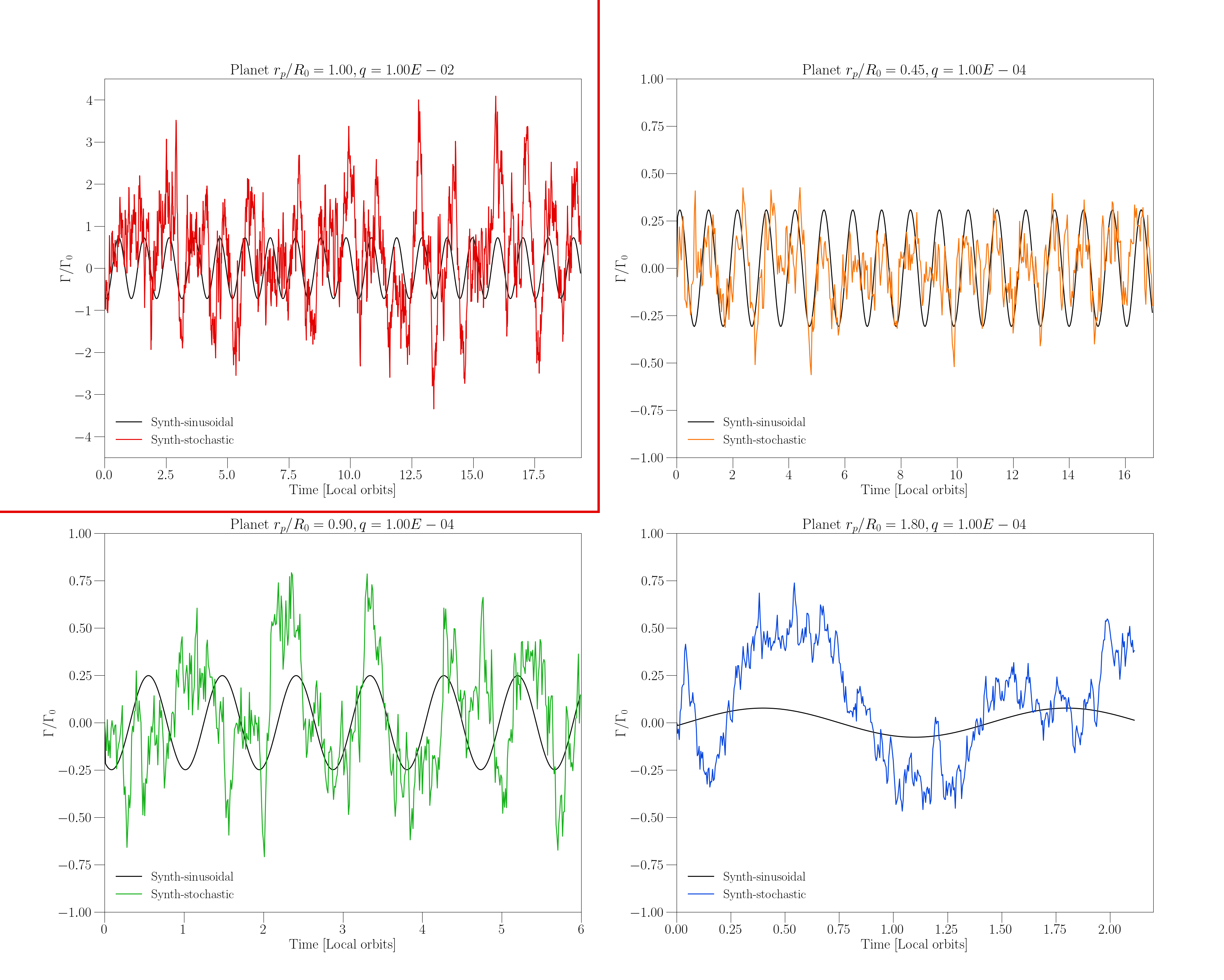}
\caption{For each planet, the total synthetic torque that produces the best-fitting autocorrelation function shown in Figure \ref{fig:autocorrelation} (color lines). The sinusoidal component embedded into the torque is overdrawn (black lines). This component has been stretched vertically ten times to highlight its features.}
\label{fig:synthetic}
\end{figure*}

Based on the measured ACRFs (color curves in Figure \ref{fig:autocorrelation}), we assume that the planetary torque can be composed of both a stochastic and a periodic component. Concerning the former, we first build a discrete first order Markov process using \(g(\Delta t)\) to generate a stochastic torque with the correlation time $\tau_c$ \citep[see][]{Kasdin_1995,Rein_2009}, in which the torque evolution at the next step, for each planet, only depends on the local RMS value, correlation time (in units of local orbital time, \(T_p\)), and the present torque. Torques at previous steps are not stored, which makes the method ``memoryless", and the mean value is zero.

The first order Markov chain is thus built from
\begin{equation}
    \Gamma(k \mathrm{d} t)/\Gamma_0 = \sqrt{\langle (\Gamma/\Gamma_0)^2 \rangle} X(k \mathrm{d} t),
\end{equation}
with
\begin{equation}
    X((k+1) \mathrm{d} t) = e^{-\frac{\mathrm{d} t}{\tau_c}} X(k \mathrm{d} t) + \sqrt{ 1 - e^{-2\frac{\mathrm{d} t}{\tau_c}} } r_{k+1},
\label{eq:markov}
\end{equation}
where \(\mathrm{d}t\) is the time interval between the \(k\) and \(k+1\) time step, \(k\) is an integer, and \(r_k\) is a random variable following a normal distribution
\begin{equation}
    r_k \rightarrow \mathcal{N}(\mu=0,\sigma=1)\,.
\label{eq:iid}
\end{equation}

The first term on the right hand side of Equation \ref{eq:markov} corresponds to an exponential decay of \(X\) with a timescale of $\tau_c$, while the second term corresponds to the excitation of new white noise \citep{Rein_2009}.

Yet, this stochastic model lacks the periodic perturbation and  its ACRFs thus behave differently to those in the simulations (shown in Figure \ref{fig:autocorrelation}). Therefore, we conjure a deterministic sine function that serves as the quasi-periodic term and add it to Equation \ref{eq:markov}. The new equation ends up looking like:
\begin{eqnarray}
    X((k+1) \mathrm{d} t) = & e^{-\frac{\mathrm{d} t}{\tau_c}} X(k \mathrm{d} t) + \sqrt{ 1 - e^{-2\frac{\mathrm{d} t}{\tau_c}} } r_{k+1} \nonumber \\
    & + \xi \sin{(2 \pi \psi (k+1) \mathrm{d} t + \delta)},
\label{eq:markov_sin}
\end{eqnarray}
where \(\xi\), \(\psi\), and \(\delta\) are amplitude, frequency (cycles per unit time), and phase-shift coefficients of the sine function.

We can thus generate the planetary torque using Equation \ref{eq:markov_sin}. To find the best fit parameters, we produce a set of synthetic torque time series with the amplitude \(\xi\) between \(10^{-3}\) and \(0.2\), the frequency \(\psi\) between \(0.5\) and \(1.5\), and the phase-shift $\delta$ between 0 and \(2 \pi\). It's important to note that having a large amplitude \(\xi \gtrsim 1\) will produce ACRFs that stay at high values for a long time before decaying, and having no phase shift will produce peaks that won't be able to align themselves with those from the simulation data.
%\todo{is this really $\xi$ or some other parameter?} - Changed to be clearer

From the produced set of possible synthetic torque time series, we calculate their respective ACRFs. We then employ the mean-square-error (MSE) method to systematically compare each synthetic autocorrelation function with its counterpart derived from the simulation; in other words, we compute the squared difference between both curves at each time lag, then select the parameter set \((\xi,\,\psi,\,\delta)\) for Equation \ref{eq:markov_sin} that minimizes the total MSE. This provides a quantitative “best-fit” criterion for matching the synthetic and simulated autocorrelations. The selected synthetic ACRFs are those shown in black curves along with the originals in Figure \ref{fig:autocorrelation}.

Additionally, in Figure \ref{fig:synthetic} we show the synthetic torque time series that correspond to the best-fitted ACRFs. The sinusoidal component of the synthetic torque is also overdrawn and enlarged.

For the particular iterations shown in this work, the RMS value \(\sqrt{ \langle (\Gamma/\Gamma_0)^2 \rangle}\) and correlation time \(\tau_c\) for each of the simulated planets' torque time series, along with the parameter set \((\xi,\,\psi,\,\delta)\) needed to reproduce their corresponding "best-fit" synthetic torque time series and respective ACRFs are shown in Table \ref{tab:coef}. Planet A.1 is the single planet in case A, while B.1, B.2 and B.3 are the three planets in case B (from closest to farthest to the star).

\begin{table}
\centering
\caption{Combination of coefficients \(\xi\), \(\psi\) and \(\delta\) for the various planets that generate the synthetic torques shown in Figure \ref{fig:synthetic}, which correspond to the autocorrelation functions shown in Figure \ref{fig:autocorrelation}. In the top row planets are denoted by their case And their position (from nearest to furthest from the star)}
\label{tab:coef}
\begin{tabular}{c|cccc}
\hline
Coeff./Planet & A.1 & B.1 & B.2 & B.3\\
\hline
\(\sqrt{ \langle (\Gamma/\Gamma_0)^2 \rangle}\) & $1.0384$ & $0.1541$ & $0.2756$ & $0.2702$\\
\(\tau_c/T_p\) & $0.1368$ & $0.1003$ & $0.0933$ & $0.1528$\\
\(\xi\) & $0.0696$ & $0.2000$ & $0.0902$ & $0.0284$\\
\(\psi\) & $0.9724$ & $0.9724$ & $1.0759$ & $0.7138$\\
\(\delta\) & $4.3329$ & $0.8674$ & $4.1163$ & $6.0656$\\
\end{tabular}
\end{table}

\subsection{Diffusion Coefficient and Timescale}

%%% Diffusion timescale

%%% CD values discussion
\begin{figure}[t!]
\centering
\includegraphics[trim=0mm 0mm 0mm 0mm, clip, width=\columnwidth]{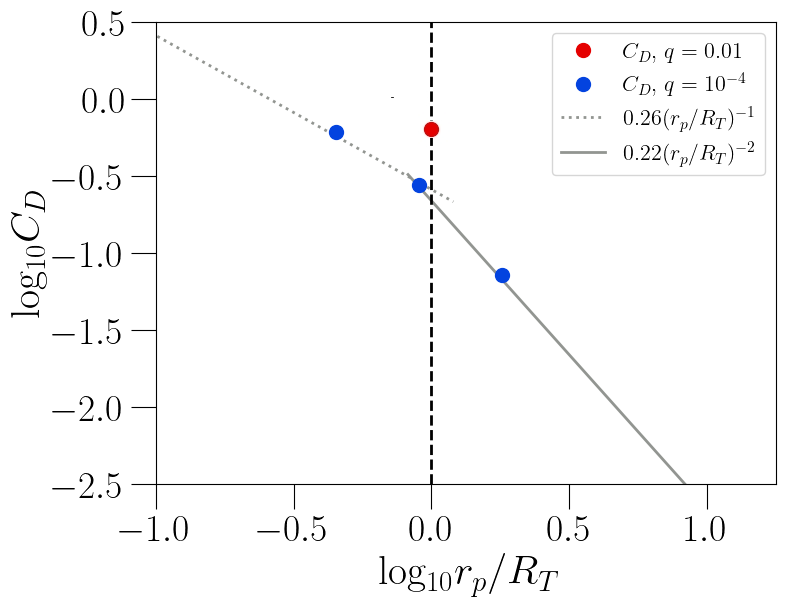}
\caption{$C_D$ values for each planet from comparing their respective planetary torque with the scaling \(2 \pi G \Sigma M_p r_p\). Possible radial dependence Equations \ref{eq:cd-rad} are shown in gray. The vertical dashed line marks the transition between the magnetosphere and turbulent disk regions (magnetosphere truncation radius).}\label{fig:cd}
\end{figure}

%%% Graph of Sigma and alpha
\begin{figure}[t!]
\centering
\includegraphics[trim=0mm 0mm 0mm 0mm, clip, width=\columnwidth]{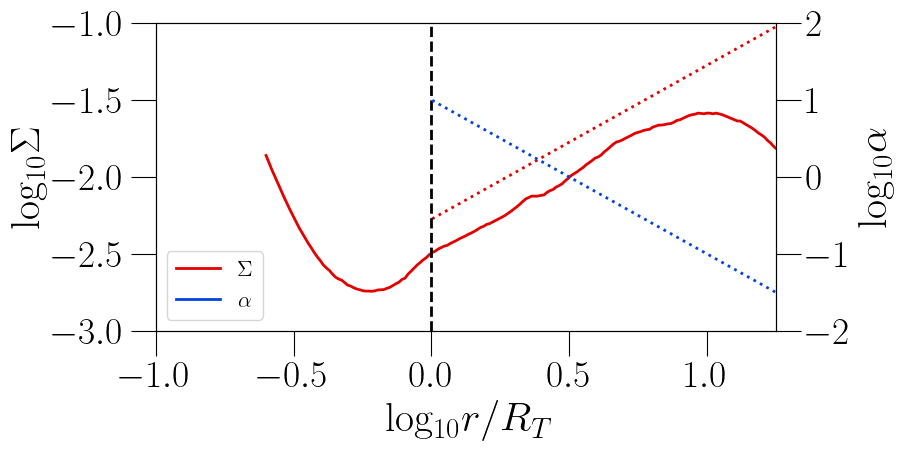}
\caption{Gas surface density \(\Sigma\) of the simulated disk (solid curve), along with the \(\alpha\) viscosity parameter profile from Equation \ref{eq:alpha_prof} and the resulting profile for \(\Sigma\) if applied (dotted curves).}\label{fig:sigma}
\end{figure}

%%% Sigma fluctuations
\begin{figure}[t!]
\centering
\includegraphics[trim=0mm 0mm 0mm 0mm, clip, width=\columnwidth]{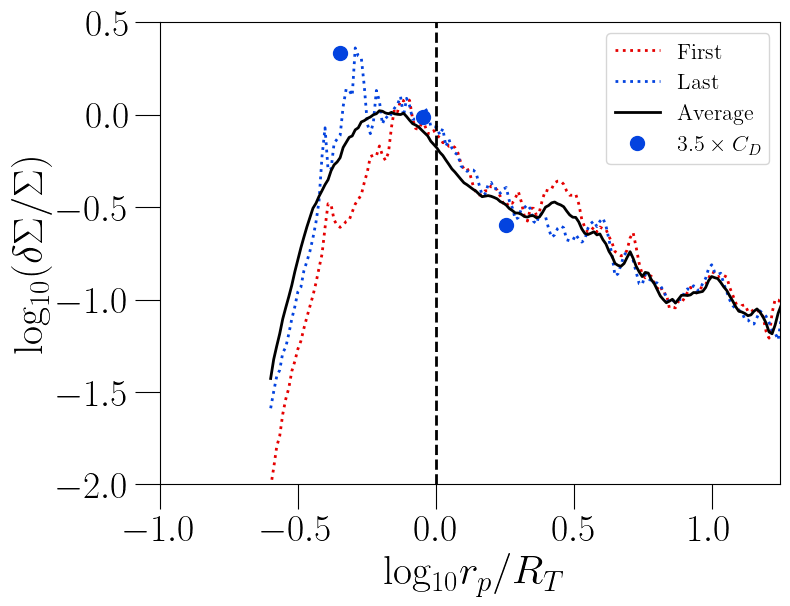}
\caption{Radial profile of the surface density fluctuation, $\log_{10}(\delta \Sigma / \Sigma)$. The red (dotted) and blue (dotted) curves represent the first and last time snapshots, respectively, while the black (solid) curve shows the time-averaged profile. The blue dots correspond to $3.5 \times C_D$, indicating a possible correlation between  $\delta \Sigma / \Sigma$ and $C_D$.}
\label{fig:deltasigma}
\end{figure}

We now compare the measured planetary torque variance with the normalization scale \(\Gamma_0\), and use Equation \ref{eq:fluctuations} to obtain average values of \(C_D\) to be $0.670$ for case A and $0.551$, $0.258$ and $0.066$ for each of the planets (from closest to farthest) for case B. These values with respect to radius are plotted in Figure \ref{fig:cd}. From this result, we presume a radial and planet mass dependence for \(C_D\). The mass dependence results from the feedback from the planet to the disk, and this dependence is relatively weak. Using the smaller planets' values, two radial relations for \(C_D\) are proposed, for both inside \(R_T\) and outside it:
\begin{equation}
    \begin{array}{rcll}
        C_D &=& 0.26 \left(\frac{r_p}{R_T}\right)^{-1}, & \text{for } r_p < R_T \\
        C_D &=& 0.22 \left(\frac{r_p}{R_T}\right)^{-2}, & \text{for } r_p > R_T\,,
    \end{array}
    \label{eq:cd-rad}
\end{equation}
which are drawn in Figure \ref{fig:cd}.

Taking the simulated disk's vertically integrated surface density, its aspect ratio radial profile $h(r=R_0)=0.1$, and the simulation's typical accretion rate of \(\dot{M}=-0.005\) (see ZSC24's section 5.4), we utilize \(\dot{M}=3 \pi \nu \Sigma\) and \(\nu = \alpha c_s H\), to estimate the simulation's \(\alpha\) parameter. We find that, for the data past the truncation radius, the profile 
\begin{equation}
    \alpha = 10 \left( \frac{r}{R_T} \right)^{-2}
\label{eq:alpha_prof}
\end{equation}
represents the data well. The simulation's \(\Sigma\) and both the \(\alpha\) (Figure \ref{eq:alpha_prof}) and complementary \(\Sigma\) profiles ($\Sigma=\dot{M}/3\pi\nu$) are shown in Figure \ref{fig:sigma}.

These relationships permit us to propose an \(\alpha\) dependence for \(C_D\) outside of the magnetosphere region. Combining Equations \ref{eq:cd-rad} and \ref{eq:alpha_prof}, we get
\begin{equation}
    C_D \approx 2\times 10^{-2}\alpha, \text{for } r_p > R_T, \label{eq:cdalpha}
\end{equation}
which further links the disk's local turbulent parameter to turbulent torque in planetary migration.

%%% Extra expression 
Additionally, by plugging $\Sigma=\dot{M}/3\pi\nu$, \(C_D\) (Equation \ref{eq:cdalpha}), and $\tau_c=0.1~T_p$ into Equation \ref{eq:diffusion0}, we derive the migration timescale due to turbulent diffusion 
\begin{equation}
t_{\mathrm{diff}}=5.6\times 10^4 \frac{h_p^4}{T_p}\left(\frac{M_*}{\dot{M}}\right)^2   \text{ for } r_p > R_T\,,
\end{equation}
where $h_p$ is the disk aspect ratio ($H/r_p$ or $c_s/v_{\phi}$) at the planet location. Please note that the turbulent diffusion timescale has no explicit dependence on $\alpha$. If $h\propto r^{1/4}$ as in our simulation, the diffusion timescale decreases quickly with respect to the  planet's distance to the star ($\propto r_p^{-1/2}$ ) when the planet is near the magnetospheric truncation radius.
%\todo{I think it should be $r_p^{-1/2}$} - DONE
% $\propto r_p^{-5/2}$ was for when we didn't have a profile for CD and assumed it to be constant

%%% About delta Sigma
Finally, Figure \ref{fig:deltasigma} presents the radial profiles of the surface density fluctuations along the azimuthal direction in the simulation, \(\delta \Sigma / \Sigma\), with scaled \(C_D\) values, where \(\delta \Sigma\) is the each annulus' surface density's standard deviation. This correlation between $\delta \Sigma/\Sigma$ and $C_D$, alongside the previous analysis, gives further evidence that regions with greater surface density perturbations, driven by disk turbulence, lead to enhanced stochastic torque variations \citep[e.g.,][]{Nelson_2004,Johnson_2006}.
%\todo{How did you define $\delta \Sigma$? Is that standard deviation?}

\section{Discussion}\label{sec:disc}

\subsection{Planet Migration: Type I/II and Aerodynamic Drag}

By scaling our simulation to a realistic protoplanetary disk that undergoes magnetospheric accretion, we are able to calculate the migration timescale of bodies embedded in the disk through various migration mechanisms. We consider a typical protoplanetary disk with an accretion rate of \(\dot{M}=10^{-8}\) \(M_{\odot}\) yr\(^{-1}\) around a Solar analogue (1 \(R_{\odot}\), 1 \(M_{\odot}\)), possessing a 1 kG dipole magnetic field.

With the code length unit \(R_0\) being set to be the truncation radius \(R_T = 10~R_{\odot} \), \(R_0\)  is thus 0.047 au. The time unit \(1/\Omega_0\) is thus 0.0016 yr. By equating the steady accretion rate of the disk \(\dot{M}= -0.005\) in code units to \(\dot{M}=10^{-8}\) \(M_{\odot}\) yr\(^{-1}\), the mass unit becomes \(3.19 \times 10^{-9}\) \(M_{\odot}\). The surface density unit is then 13.1 g cm$^{-2}$.

Figure \ref{fig:sigma} thus suggests that \(\Sigma \sim 0.2\) g/cm\(^2\) at \(r \sim 0.2\) au. This low surface density is due to the large \(\alpha\) value from the strong turbulence. As concluded by ZSC24, even with more realistic aspect ratios making the surface density \(\sim 10\) times larger and dust being a small percentage of the total mass, the resulting mass of solids is orders of magnitude less than what is required to explain the abundant exoplanets close to the star. The MRI-active inner disk appears to be highly unfavorable for planet formation, indicating that planets are more likely to form in the more massive deadzone. 

%%% Making a more realistic disk
To study the migration timescales, we follow \cite{Zhu_2025} to assume a realistic disk aspect ratio as \(h = 0.05 (r/R_T)^{1/4}\). Additionally, for a more realistic disk that is in the spin equilibrium state, we adopt 
\begin{equation}
\alpha =\frac{R_T}{r}\,,     \label{eq:alphap}
\end{equation}
as shown in \cite{Zhu_2025}.
Thus
\begin{equation}
    %\Sigma = \frac{\dot{M}}{3\pi}(GM_*r)^{-1/2}(\alpha h^2)^{-1} = K (GM_*r)^{-1/2} \left( \frac{r}{R_T} \right)^{1/2}
    \Sigma(r) = \frac{\dot{M}}{3\pi}(GM_*r)^{-1/2}(\alpha h^2)^{-1} \approx 2.78 \text{ g cm}^{-2},
\end{equation}
which is coincidentally independent from the disk radius.

%%% Migration timescale heat map review
\begin{figure}[t!]
\centering
\includegraphics[trim=0mm 0mm 0mm 0mm, clip, width=\columnwidth]{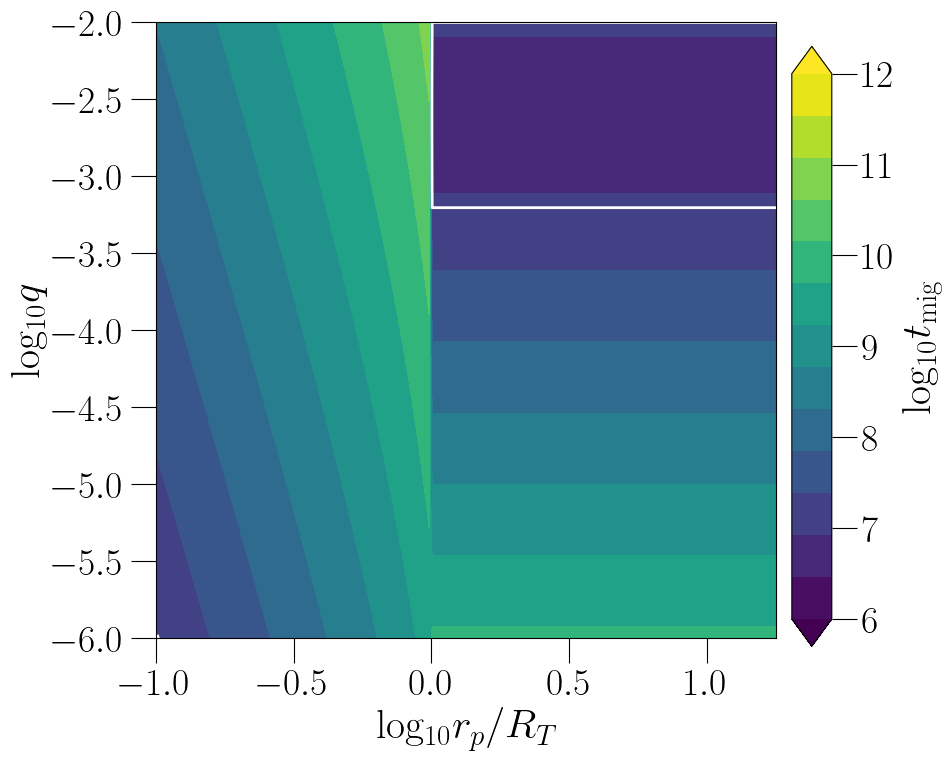}
\caption{Migration timescales from Type I/II migration and aerodynamic drag for a planet in the inner disk. The magnetospheric truncation radius $R_T$, equivalent to 10 solar radii, divides the graph into two regions. To the right lies the turbulent disk, where Type I/II timescales are calculated (Equation \ref{eq:disk-driven}), while to the left lies the magnetosphere region, where the planet is instead subjected to aerodynamic drag (Equation \ref{eq:drag}). The solid white contour labels the migration timescale of \(10^7\) yrs. The central star is a solar analogue.}\label{fig:heatmap}
\end{figure}

On the other hand, for any planet that somehow manages to go through the inner MRI turbulent disk and enter into the magnetosphere, the planet cannot interact with the disk anymore (Equation \ref{eq:disk-driven}  becomes invalid). Instead, the planet would be subjected to strong aerodynamic drag and continue its migration towards the central star if the planet is inside the corotation radius. From ZSC24, the planet's migration rate through aerodynamic drag inside the inner cavity is governed by
\begin{equation}
    t_{\mathrm{drag}} = \frac{4 R_p \rho_p}{3 v_{\mathrm{rel}} \rho},
\label{eq:drag}
\end{equation}
where $R_p$ and $\rho_p$ are the planet's radius and its material density (fixed to 3 g cm$^{-3}$), respectively. \(v_{\mathrm{rel}} = v_K - \Omega_{*}r_p\) is the relative velocity between the planet and the background magnetosphere plasma with \(\Omega_{*}\) being the star's spin rate. Assuming that the star is in the spin-equilibrium state with the disk, we have \(R_T = 0.79 R_C\) \citep{Zhu_2025}, where \(R_C\) is the corotation radius. With \(R_C\) given, we can compute \(v_{\mathrm{rel}}\). This is different from ZSC24, where the non-rotating star leads to \(v_{\mathrm{rel}} \sim v_K\).

With this disk model, we calculate the migration timescale for planets at different distances from the star through Type I/II migration (Equation \ref{eq:disk-driven}) or aerodynamic drag (Equation \ref{eq:drag}), as illustrated in Figure \ref{fig:heatmap}. Given that the lifetime of protoplanetary disks is estimated to be between \(<1\) and 40 Myr \citep[see e.g.,][]{Pfalzner_2024}, we find that for most low-mass planets the disk-driven migration timescale exceeds the disk's lifetime. Only the massive planets are able to reach the magnetospheric truncation radius, with planets of mass ratio \(q\approx0.003\) being the most favored. This implies that, within the inner MRI-active region, significant migration is unlikely for most planets except at the very early times when the accretion rate or surface density is much higher. Once a planet reaches the truncation radius, it enters the aerodynamic drag regime, where its migration timescale increases by several orders of magnitude. With no additional torques, the massive planets become effectively trapped at the truncation radius. For gas giants, this truncation radius may represent the innermost limit of their migration \citep[see e.g.,][]{Mendigutia_2024}.

Since low mass planets cannot migrate in the inner MRI active disk, they are likely to stay at the deadzone inner boundary (DZIB). 
Before migrating planets even enter the MRI-active region (active when \(T \gtrsim 1000\) K), they must first pass through the deadzone inner boundary. The DZIB marks the transition between the inner, sufficiently ionized, highly turbulent “MRI-active” zone and the outer, much more massive, low-ionization, less turbulent “MRI-dead” zone. This abrupt transition can create a jump in the surface density, which, in turn, generates a positive corotation torque. Such a torque may be strong enough to act as a pebble trap \citep[see e.g.,][]{Chatterjee_2014} and potentially even function as a planet trap, particularly affecting intermediate-mass planets in the range of \(q \sim 10^{-5}\text{ --- }10^{-4}\) \citep[e.g.,][]{Hu2016, Faure_2016}. Thus, low mass planets could form or be trapped at the DZIB.
For T Tauri stars, the DZIB is located at $\sim$0.1 au \citep[e.g.,][]{dalessio1998,Flock_2019}, which is slightly larger than the magnetospheric truncation radius. For Herbig Ae/Be stars, the DZIB can extend to 1 au \citep[e.g.,][]{dullemond2010}, which is significantly larger than the truncation radius, especially considering Herbig Ae/Be stars have much weaker magnetic fields \citep[see e.g.,][]{Jarvinen_2019}. This implies that low mass planets around Herbig Ae/Be stars should be further out in the disk.

\subsection{Planet migration: Turbulent Diffusion}

%%% Talking about timescale figure comparing migration and diffusion
\begin{figure}[t!]
\centering
\includegraphics[trim=0mm 0mm 0mm 0mm, clip, width=\columnwidth]{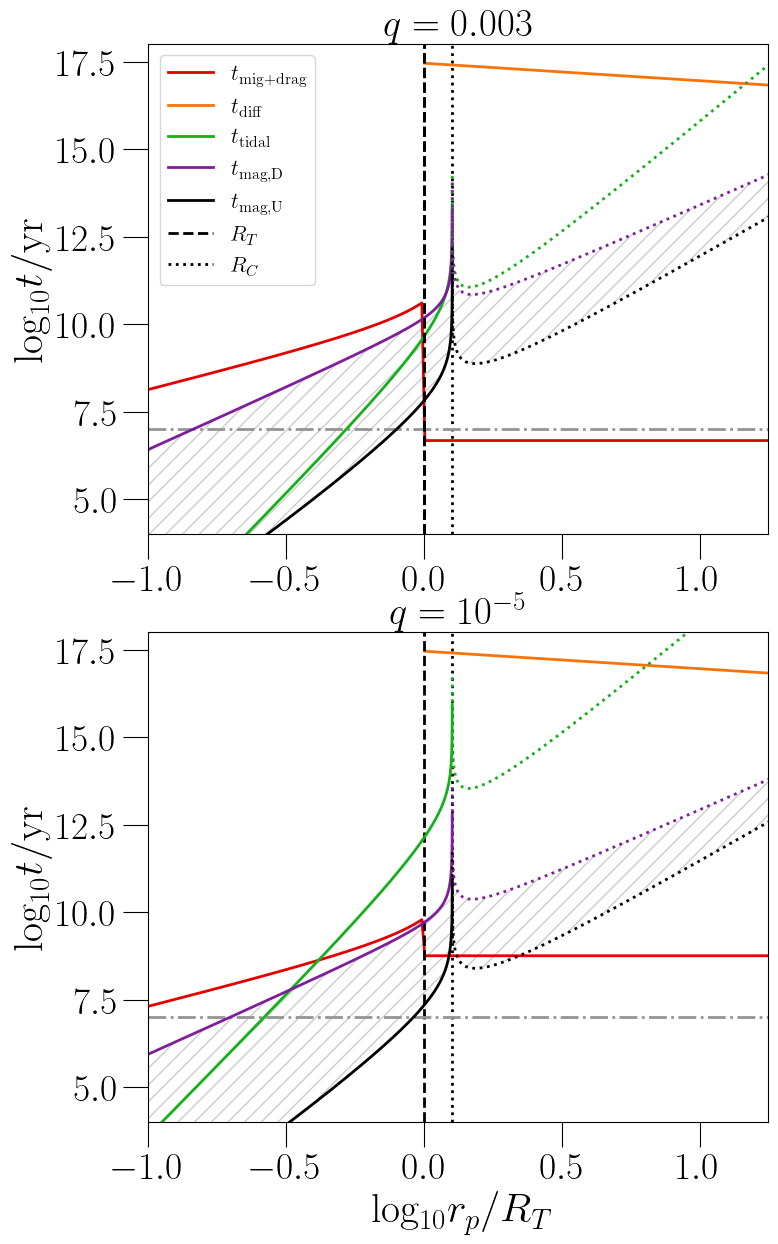}
\caption{Comparison of migration timescales for two representative planet-star mass ratios: \(q=0.003\) (top) and \(q=10^{-5}\) (bottom).
The disk-driven plus drag timescale is in red (Equations \ref{eq:disk-driven} and \ref{eq:drag}), while diffusion is in orange (Equation \ref{eq:diffusion-new}). Tidal migration appears in green (Equation \ref{eq:tidal_time}). Two magnetic torque regimes, dipolar (\(t_{\mathrm{mag,D}}\), purple) and unipolar (\(t_{\mathrm{mag,U}}\), black), follow Equation \ref{eq:mag_time} and \ref{eq:mag_time_unipol}, respectively.
The region between these two magnetic curves is hatched to indicate the range of possible magnetic migration timescales, reflecting uncertainties in planetary field strength, conductivity, stellar wind properties, angle, etc.
Solid curves denote inward migration, whereas dotted ones denote outward migration. The dash-dotted horizontal line at \(10^7\)\,yr marks a typical disk lifetime, and the vertical dashed/dotted lines indicate radii of interest (\(R_T\) and \(R_C\), respectively).}
\label{fig:timescale}
\end{figure}

Our ACRF analysis in Section \ref{sec:results} suggests that the correlation timescale \(\tau_c\) is shorter than estimates in other works \citep[see e.g.,][]{Johnson_2006,Adams_Bloch_2009}, where values of half or a full-orbit are suggested; instead, we measure values in order of \(\sim 0.1 \:T_p\). This means that Equation \ref{eq:diffusion} needs to be increased by a factor of 5 to:
\begin{equation}
    t_{\mathrm{diff}} = \frac{5}{4 \pi^3 C_D^2 \bar{\Sigma}^2} \left( \frac{M_{\star}^3}{G r_p^5} \right)^{1/2}\,.
\label{eq:diffusion-new}
\end{equation}

We compare the turbulent diffusion timescale of Equation \ref{eq:diffusion-new} with that of gas-driven Type I/II migration (Equation \ref{eq:disk-driven}) and aerodynamic drag (Equation \ref{eq:drag}) for two different planet-star mass ratios, \(q=0.003\) and \(q=10^{-5}\), in Figure \ref{fig:timescale}. These two $q$ values are representative of relatively quick and slow Type I migration rates, respectively. As for \(C_D\), instead of a fixed value we utilize the radial profile of Equation \ref{eq:cdalpha}. We find that, regardless of the precise value of $C_D$ or $\tau_c$ utilized, the turbulent diffusion timescale ends up being several orders of magnitude larger than that of gas-driven migration, being larger than the Hubble time in the whole region of interest. This is due to the small disk scale height and the small disk surface density from large $\alpha$. The ratio between turbulent diffusion timescale (Equation \ref{eq:diffusion0}) and the Type I migration timescale (Equation \ref{eq:disk-driven}) is
\begin{equation}
\frac{t_{\mathrm{diff}}}{t_{\mathrm{mig},I}}=\frac{1}{8\pi^3 C_D^2}\frac{M_p}{\Sigma H^2}\frac{T_p }{\tau_c}\,.
\end{equation}
With $M_p=M_{\oplus}$, $\Sigma\sim 2$ g/cm$^2$, $C_D\sim 0.02$, $h\sim 0.05$ at 0.1 au, and $\tau_c=0.1 T_p$, this ratio is 5$\times 10^7$, indicating turbulent diffusion is negligible for planets at the inner MRI turbulent disk.

As we move outwards, from the MRI active region into the deadzone, the ionization fraction decreases, prompting a decoupling of the magnetic fields from the gas. This leads to low or negligible turbulence, and the \(\alpha\) parameter could settle to a much smaller value (e.g. \(\alpha \sim 10^{-4}\)). Turbulence will then be constrained to the active layers above and below the midplane \citep[e.g.,][]{Oishi_2007}. This would result on the turbulent diffusion timescale at the midplane staying large or even increasing. Therefore, we can generally presume that stochastic torques have a negligible effect on changing a planet's semi-major axis within several au.

\begin{table*}[t!]
\centering
\caption{Parameters to evaluate tidal and magnetic torques and produce Figure \ref{fig:timescale}.}
\setlength{\tabcolsep}{14pt} % Adjust column separation to help center the table
\begin{tabular}{cccccccc}
\hline
$M_\star$ ($\odot$) & $R_\star$ ($\odot$) & $P_\star$ (days) & $B_\star$ (G) & $\tau_\star$ (s) & $M_p$ ($J$) & $R_p$ ($J$) & $B_p$ (G) \\
\hline
$1$ & $1$ & 5.2 & 1000 & 1 & $3$ --- $0.01$ & $1$ --- $0.1$ & 100 \\
\hline
\end{tabular}
\begin{flushleft}
\textbf{Note.} Adapted from \cite{wei2024magneticfieldgasgiant}. High and low mass planets' characteristics are separated by "---".
\end{flushleft}
\label{tab:parameters}
\end{table*}

\subsection{Planet Migration: Tidal Torques}

In addition to the Type I/II migration and turbulent diffusion driven by planet-disk interaction, a planet can also migrate through tidal interaction with the central star.  Figure \ref{fig:timescale} also displays the corresponding timescales for tidal and magnetic torques. Following \cite{wei2024magneticfieldgasgiant}, we use a standard tidal time-lag formalism \citep[see e.g.,][]{Hut_1981,Eggleton_1998,Ogilvie_Lin_2004,Ogilvie_Lin_2007} to express the torques due to star-on-planet and planet-on-star interactions
\begin{equation}
    \Gamma^t_{\star p} \approx G M_p^2 R_{\star}^5 (\omega_p - \Omega_{\star}) \tau_{\star}/r_p^6,
    \label{eq:tidal}
\end{equation}
\begin{equation}
    \Gamma^t_{p \star} \approx G M_{\star}^2 R_p^5 (\omega_p - \Omega_p) \tau_{p}/r_p^6,
    \label{eq:tidal-planet}
\end{equation}
where the tidal time lags $\tau_{\star}$ and $\tau_p$ represent the delay between the tidal forcing and the tidal response in the star and planet, respectively.

Owing to the substantial mass difference between the star and the planets, and their short orbital separations, planets would typically synchronize their spin ($\omega_p \approx \Omega_p$) on timescales of a few million years. Consequently, the star-on-planet tidal torque in Equation \ref{eq:tidal-planet} quickly becomes negligible before significant migration takes place, leaving the planet-on-star torque of Equation \ref{eq:tidal} dominant.

% Q-value first part
The tidal time lag is directly related to the tidal quality factor ($Q$), defined as $Q \simeq (\omega \tau)^{-1}$, where $\omega$ is the tidal frequency. For close-in planets, a representative tidal frequency is $\omega \sim 10^{-5}\mathrm{s}^{-1}$. Higher values of $Q$ indicate less efficient tidal dissipation and slower orbital evolution, while lower values imply stronger dissipation and more rapid migration. T Tauri stars generally have $Q \sim 10^4$, but as they evolve onto the main sequence, becoming less convective and slowing their rotation, $Q$-values increase significantly, diminishing tidal effects \citep{Gallet_2017}. We choose a stellar tidal time lag of $\tau_{\star}=1 s$, which is equivalent to the
stellar tidal quality factor of $Q=10^5$ for 7 days periods.

Young stars rotate fast and our assumed stellar rotation period of 5.2 days, using \(P_* = 2\pi\sqrt{R_C^3/(GM_*)}\), corresponds to a corotation radius $R_C$ of roughly $0.06\,$au, near the disk truncation radius. For a planet interior to this radius, its orbital motion is faster than the star's rotation. Thus,
the stellar tide provides a negative torque to the planet's orbital motion, leading to the planet's inward migration. On the contrary, for a planet outside the corotation radius, it migrates out. Thus, the tidal migration is a divergent migration process further away from the corotation radius. 

For a planet on a circular orbit being subjected to a torque $\Gamma$, its semimajor axis evolves at a rate
\begin{equation}
    \dot{r_p} = 2 \Gamma \frac{r_p^{1/2}}{M_p \sqrt{GM_{\star}}}.
    \label{eq:rate}
\end{equation}
Thus, using Equations \ref{eq:tidal} and \ref{eq:rate} with the definition of the migration timescale \(r_p/|\dot{r_p}|\), we obtain the tidal migration timescale
\begin{equation}
    t_{\mathrm{tidal}} = \frac{1}{2}R_{\star}^{-5}\tau_{\star}^{-1}|\omega_p - \Omega_{\star}|^{-1} q^{-1} \left( \frac{r_p^{13}}{GM_{\star}} \right)^{1/2}.
    \label{eq:tidal_time}
\end{equation}

\subsection{Planet Migration: Magnetic Torques}

Short-period planets are subject to magnetic interactions with the stellar magnetic fields, which can also generate substantial torques on the planets. The motion of a planet through the stellar magnetosphere produces Alfvén wings, which transport energy and angular momentum along magnetic field lines \citep[see, e.g,][]{Neubauer_1998,Saur_2013}. Depending on the stellar wind parameters, field geometry, and the planet’s own properties, one or both of these Alfvén wings may connect back to the star \citep[see][]{Strugarek_2015}, setting the overall coupling strength.

A key discriminator is whether Alfvén waves can make a round trip between the planet and star faster than magnetic field lines “slip” around the planet \citep[][]{Saur_2013, Strugarek_2015, Strugarek_2016}. If the round-trip transit time for Alfvén waves is longer than the slipping time, the interaction is often labeled dipolar because the planetary dipole (if present) dominantly opposes or redirects the stellar field, but no fully closed circuit forms back to the star. If, however, the planet’s conductivity and geometry allow Alfvén waves to travel back and forth before the field lines slip away, the system enters the unipolar regime. This latter case leads to stronger electrical coupling (akin to the Io–Jupiter circuit) and can yield large torques \citep[see][]{Laine_2008, Laine_2012, Strugarek_2017}.
% and planet inflation due to ohmic dissipation

The dipolar regime is expected for most compact star–planet exosystems \citep[][]{Strugarek_2017}, especially when the planet generates its own magnetosphere against the ambient stellar wind or magnetized plasma, since the Alfvén waves wouldn't have the time to do the round-trip to establish an enduring circuit. Alternatively, if the planet is weakly magnetized or unmagnetized, but has high magnetic diffusivity, the stellar field simply penetrates and is dissipated, effectively decoupling or “slipping” from the planet.

In these cases, time-dependent perturbations in the stellar magnetic field typically do not form a full conduction loop back to the star. Instead, the steady (or slowly varying) component of the stellar field might be partially screened in the planet or compressed against its magnetopause. The net effect is a magnetic torque that can, nonetheless, drive planetary spin–orbit evolution, but typically at a much weaker level than in the unipolar regime \citep[][]{Strugarek_2017}.

When the planet is able to sustain its own magnetosphere, \cite{wei2024magneticfieldgasgiant} approximates the star-on-planet magnetic torque $\Gamma^m_{\star p}$ from the dipolar interaction as
\begin{equation}
    \Gamma^m_{\star p} \approx 4 (\omega_p - \Omega_{\star}) \left( \frac{B_{\star}^2}{\Lambda} \right) r_p^2 R_{\mathrm{obs}}^2 \left( \frac{R_{\star}}{r_p} \right)^6,
    \label{eq:magtorque_robs}
\end{equation}
where $\Lambda$ denotes the effective Alfvén-wing resistance, and \(R_{\mathrm{obs}}\) is the planet's obstacle radius. For a magnetized planet, they derive \(R_{\mathrm{obs}} \approx R_p (B_p/B_{\star})^{1/3}(r_p/R_{\star})\), where $B_{\star}$ and $B_p$ are the respective stellar and planetary magnetic field strengths at their equators. Thus, the planet's dipole boosts the effective interaction cross section, giving
\begin{equation}
    \Gamma^m_{\star p} \approx 4 (\omega_p - \Omega_{\star}) \left( \frac{B_{\star}^2}{\Lambda} \right) r_p^2 R_p^2 \left( \frac{B_p}{B_{\star}} \right)^{2/3} \left( \frac{R_{\star}}{r_p} \right)^4.
    \label{eq:magtorque}
\end{equation}

From this, the characteristic orbital migration due to dipolar magnetic interaction can be defined as
\begin{eqnarray}
    t_{\mathrm{mag, D}} &= \frac{1}{8} | \omega_p - \Omega_{\star}|^{-1} \left( \frac{\Lambda}{B_{\star}^2} \right) r_p^{-3/2} R_p^{-2} \left( \frac{B_{\star}}{B_p}\right)^{2/3} \nonumber \\
    &\times \left( \frac{r_p}{R_{\star}} \right)^4 M_p (G M_{\star})^{1/2}.
    \label{eq:mag_time}
\end{eqnarray}

By contrast, in the unipolar regime when the Alfvén waves traveling along the magnetic fields settle into a closed circuit before the field lines slip away, the planet effectively drags the stellar magnetic field lines. This is particularly relevant for low planetary diffusivities, so that the stellar field “freezes into” the planet’s interior (rather than being dissipated), and for planets that are weakly magnetized or unmagnetized but nonetheless highly conducting. The conduction path extends through the star’s envelope at the field-line footpoints, giving rise to a complete circuit \citep[see e.g.,][]{Goldreich_1969, Laine_2008}.

This is thoroughly explained in \citet{Laine_2012}, where they consider a planet moving through the stellar magnetic fields and the planet has negligible intrinsic fields. A voltage is induced across the planet, which drives a current through the flux tube connecting the planet and the star. The circuit closes primarily through the stellar envelope, where the largest resistance typically resides. In this picture, the effective obstacle radius is simply the planet’s physical cross section, \(R_{\mathrm{obs}} \equiv R_p\). The unipolar net torque is then approximated by
\begin{equation}
    %\Gamma^m_{\star p} \approx 16 (\omega_p - \Omega_{\star}) r_p^2 R_p^2 \left( \frac{\mu_0 m}{4 \pi r_p^3} \right)^2 \Sigma_v s,
    \Gamma^m_{\star p} \approx 16 (\omega_p - \Omega_{\star}) r_p^2 R_p^2 \left( \frac{m}{r_p^3} \right)^2 \Sigma_v s\,\:{\rm in \: C.G.S.,}
\end{equation}
where \(m\) is the stellar dipole moment (related to \(B_{\star}\) by \(m \simeq B_{\star} R_{\star}^3\)), and \(\Sigma_v\) the vertical conductance in the stellar envelope. The factor \(s = \cos(\theta_F) = \sqrt{1-R_{\star}/r_p}\), where \(\theta_F\) is the angle between the stellar spin axis and the location of the foot of the flux tube. If we assume them to be nearly aligned, we have \(s \approx 1\). Additionally, the total resistance of the stellar atmosphere at the foot of the flux tube is given by \(\mathcal{R}_{\star}=(2\Sigma_vs)^{-1}\). Thus, we can rewrite the previous equation as
\begin{equation}
    \Gamma^m_{\star p} \approx 8 (\omega_p - \Omega_{\star}) \left( \frac{B_{\star}^2}{\mathcal{R}_{\star}} \right) r_p^2 R_p^2 \left( \frac{R_{\star}}{r_p} \right)^6,
    \label{eq:unipol_torque}
\end{equation}
and the associated characteristic unipolar timescale as
\begin{eqnarray}
    t_{\mathrm{mag, U}} &= \frac{1}{16} | \omega_p - \Omega_{\star}|^{-1} \left( \frac{\mathcal{R}_{\star}}{B_{\star}^2} \right) r_p^{-3/2} R_p^{-2} \nonumber \\
    &\times \left( \frac{r_p}{R_{\star}} \right)^6 M_p (G M_{\star})^{1/2}.
    \label{eq:mag_time_unipol}
\end{eqnarray}

We assume Equations \ref{eq:unipol_torque} and \ref{eq:mag_time_unipol} apply to all planets in our sample, with the caveat that they strictly hold only if the resistance at the star’s footpoint dominates over that inside the planet. This approximation was originally derived for super-Earths, where interior conductivity is expected to be high, but still less than in the stellar atmosphere. By contrast, gaseous giants might have comparable or even higher interior conductivity, but this is currently an area that remains poorly constrained \citep[see e.g.,][]{Lai_2012,Strugarek_2017}.

Numerically, \cite{Laine_2012} find \(\mathcal{R}_{\star} \approx 8.6 \times 10^{-6}\) ohm at the stellar envelope footpoint. Meanwhile, the Alfvén-wing resistance can be approximated by \(\Lambda \approx 0.25 |\omega_p - \Omega_{\star}|r_p/(10^7 \mathrm{cm \, s}^{-1}) + 0.1\) ohm \citep[][]{Lai_2012,Laine_2008}. Using our T Tauri parameters, we find \(\Lambda/\mathcal{R_{\star}}\) can range from $\sim 10^3$ to $10^5$. Consequently, based on Equations \ref{eq:magtorque_robs} and \ref{eq:unipol_torque}, the unipolar interaction can produce torques orders of magnitude larger than the dipolar case, leading to substantially shorter magnetic migration timescales. 

\subsection{Implications for planet demographics}

Figure \ref{fig:timescale} groups together the timescales for all the migration pathways previously discussed (Equations \ref{eq:disk-driven}, \ref{eq:drag}, \ref{eq:diffusion-new}, \ref{eq:tidal_time}, \ref{eq:mag_time}, \ref{eq:mag_time_unipol}). 
Table \ref{tab:parameters} shows the parameters used to calculate the tidal and magnetic torques for our chosen mass ratios, $q=0.003$ and $q=10^{-5}$. We assume a plausible range of planetary radii (\(R_p = 1~M_J\) vs.\ \(R_p = 0.1~M_J\)). The stellar rotation period is fixed at 5.2 days so we have $R_C = 0.06$ au and set the truncation radius at \(R_T = 0.79~R_C\). We also take $B_p/B_{\star} = 0.1$ in both scenarios. We take Equations \ref{eq:mag_time} and \ref{eq:mag_time_unipol} to serve as an upper and lower limit of the magnetic torque, respectively, with a shaded region between them.

\begin{figure*}[t!]
\centering
\includegraphics[trim=0mm 0mm 0mm 0mm, clip, width=7.in]{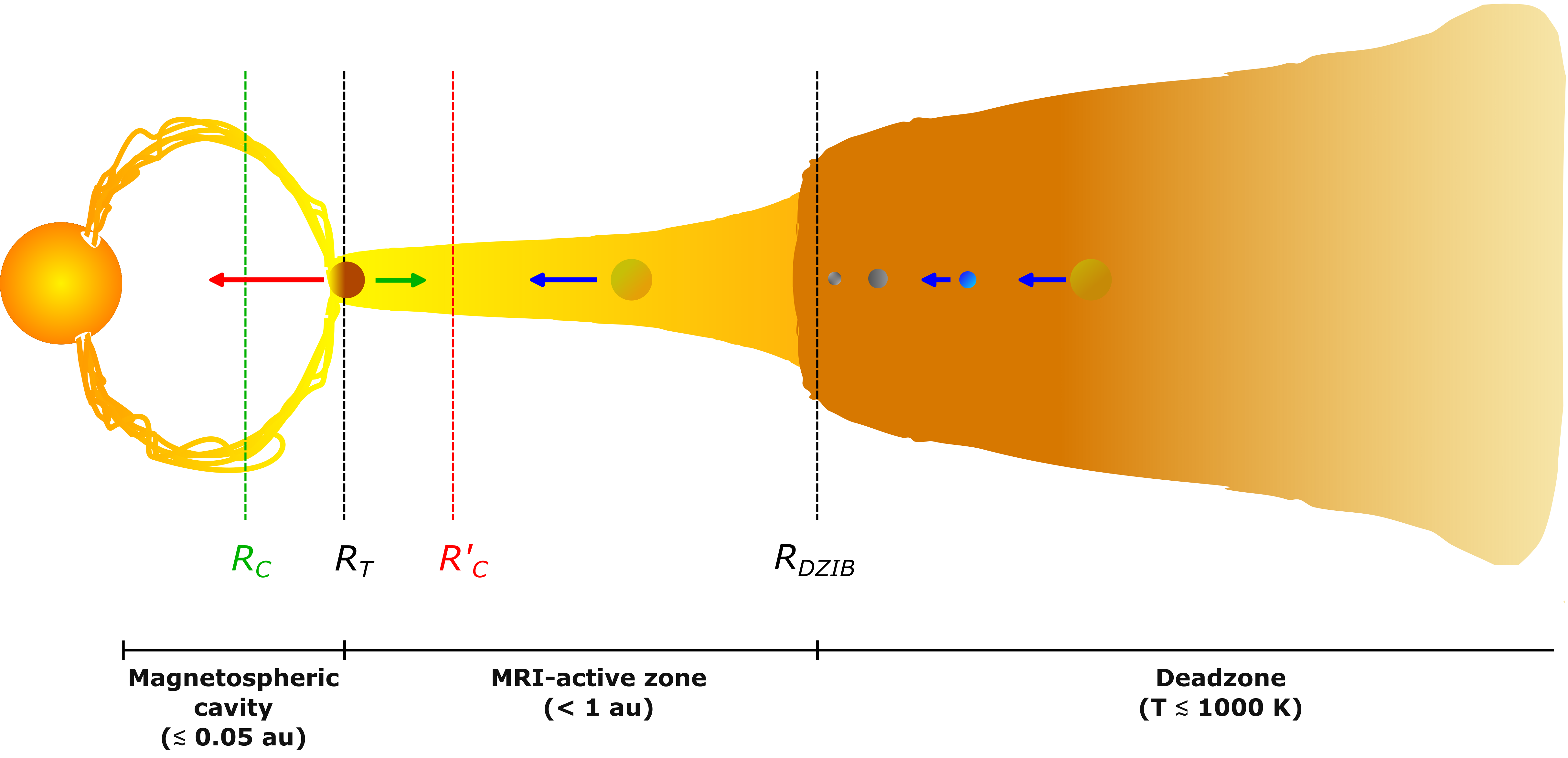}
\caption{Schematic depiction of possible end states for migrating planets. In the low-mass regime, planets tend to stall near the deadzone inner boundary (DZIB). More massive planets can move inward across this boundary and reach the MRI-active zone. If they encounter the magnetospheric truncation radius ($R_T$), the planet’s fate then hinges on the location of the stellar corotation radius ($R_C$). If $R_T > R_C$, planets remain stranded at $R_T$. Conversely, if $R_T < R_C$, they can be drawn into the magnetosphere and eventually fall onto the star.}\label{fig:diagram}
\end{figure*}

Figure \ref{fig:timescale} presents various migration timescales at the inner MRI active disk, showing that disk-driven migration (Type I/II) dominates at the disk midplane. Jupiter-like planets migrate faster than rocky planets since $t_{mig}$ is inversely proportional to $q$ when a gap is not induced (Equation \ref{eq:disk-driven}). With the strong turbulence ($\alpha\sim$1), even a Jupiter mass planet fails to induce a gap. Once a giant planet enters the magnetospheric cavity, disk-driven migration becomes negligible and the aerodynamic drag between the magnetosphere and the planet becomes more important, whose migration timescale increases to roughly the lifetime of a solar-type star. This effectively “traps” the gas giant at or near the truncation boundary. 
Tidal and magnetic torques start to dominate the migration timescale at such a distance to the central star.
Especially, as the planet moves closer to the star, these forces will grow significantly. For instance, for $q=0.003$, the tidal torque timescale at $R_T$ remains longer than disk-driven migration but is still a fraction (around one-tenth) of the star’s main-sequence lifetime. Although magnetic torques seem to be the most efficient, they are uncertain depending on both stellar fields and planet conductivity.
The stellar fields typically weaken on a time scale of tens of millions of years.
%\todo{What is $R_0$ in the figure?} - DONE. Makes more sense to have $R_T$ in the graphs instead of $R_0$.

In this setup, a $q=0.003$ planet experiences enough tidal and magnetic torque to eventually free it from $R_T$, causing it to spiral inward more quickly. Consequently, if a massive planet undergoing Type I/II migration arrives at $R_T$ before disk dissipation and if $R_C$ lies beyond $R_T$ ($R_C > R_T$), these forces can pull the planet into the magnetosphere, leading to orbital decay and eventual infall to the star. Alternatively, if $R_C < R_T$, the planet remains trapped near $R_T$ through the combination of drag and the outward push of tidal and magnetic torques.

On the other hand, for $q=10^{-5}$, the inward migration due to disk-driven torques is so slow that the planet is unlikely to reach $R_T$ before the disk dissipates. Although migration due to magnetic interaction is faster with a smaller mass planet, it is still too long compared with the disk lifetime beyond the magnetic truncation radius. 

Recent planet demographic works by \cite{Mendigutia_2024} and \cite{Sun_2025} indicate that gas giants are commonly discovered near the magnetospheric truncation radius, while super-Earths and sub-Neptunes are predominantly curtailed by the dust sublimation radius. Our results are consistent with this trend: hot Jupiters reaching the magnetosphere can either remain at $R_T$ due to a net outward torque or instead undergo rapid inward migration, contingent upon the relative positioning of $R_C$. Figure \ref{fig:diagram} summarizes these possible pathways and emphasizes the crucial roles of disk structure, planet mass, and stellar torques in shaping system architecture.

%%% Planet population figure
\begin{figure*}[t!]
\centering
\includegraphics[trim=0mm 0mm 0mm 0mm, clip, width=7.in]{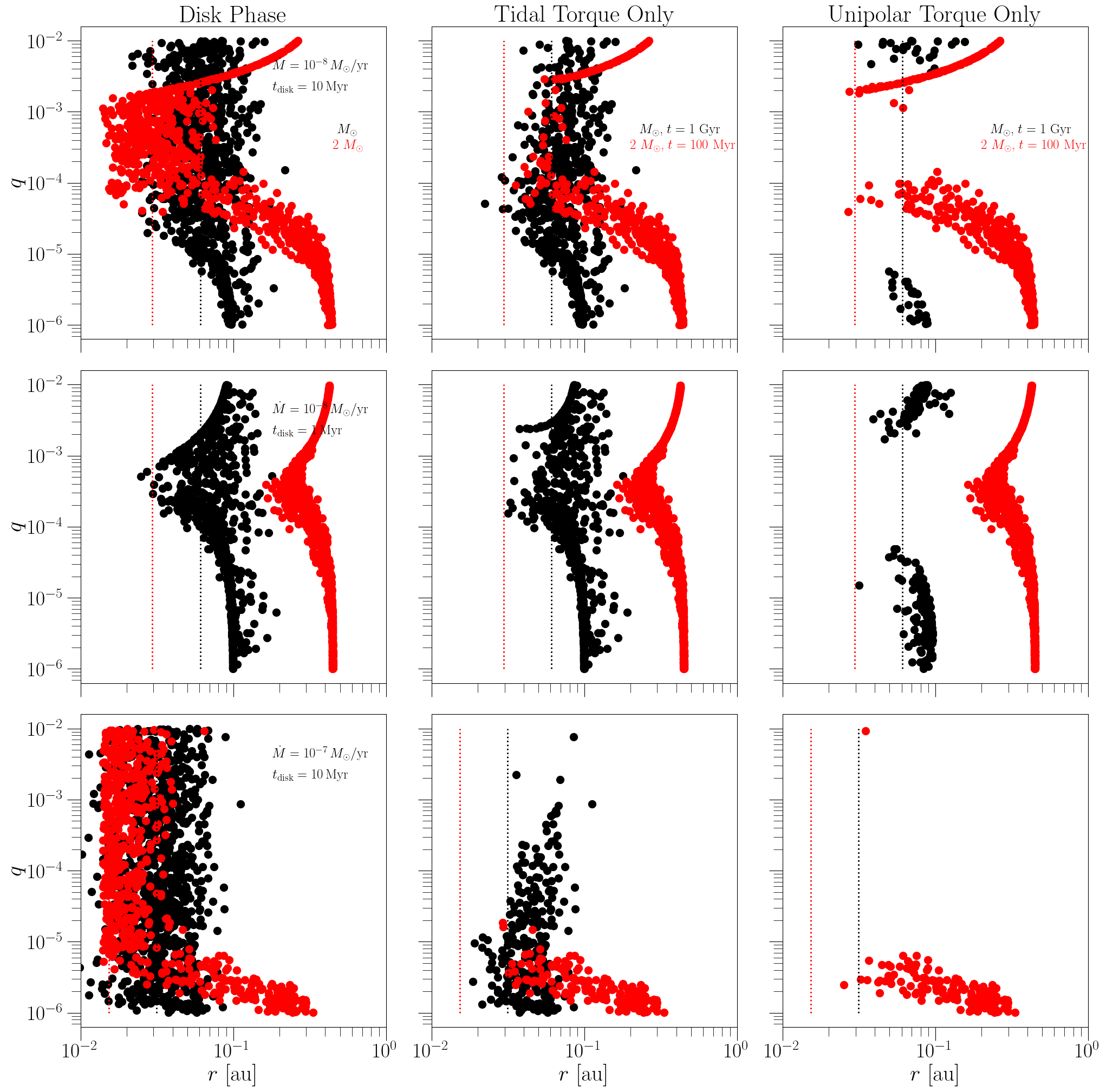}
\caption{Population synthesis results showing planetary positions after different evolutionary phases for stellar masses of 1~$M_{\odot}$ (black points) and 2~$M_{\odot}$ (red points). Each row represents a distinct disk condition: fiducial disk phase ($\dot{M}=10^{-8}\,M_{\odot}/\mathrm{yr}$, $t_{\mathrm{disk}}=10$~Myr, top row), shortened disk lifetime ($\dot{M}=10^{-8}\,M_{\odot}/\mathrm{yr}$, $t_{\mathrm{disk}}=1$~Myr, middle row), and higher disk accretion rate ($\dot{M}=10^{-7}\,M_{\odot}/\mathrm{yr}$, $t_{\mathrm{disk}}=10$~Myr, bottom row). 
The first column illustrates planet positions immediately after the protoplanetary disk dispersal. Subsequent columns show planetary distributions after an extended period of either purely tidal torque-driven migration or purely unipolar magnetic torque-driven migration. Vertical dotted lines represent the magnetospheric truncation radius calculated for each stellar mass using the initial mean stellar magnetic field strength.}
\label{fig:planet_pop}
\end{figure*}

%%% Planet fraction figure
\begin{figure*}[t!]
\centering
\includegraphics[trim=0mm 0mm 0mm 0mm, clip, width=7.in]{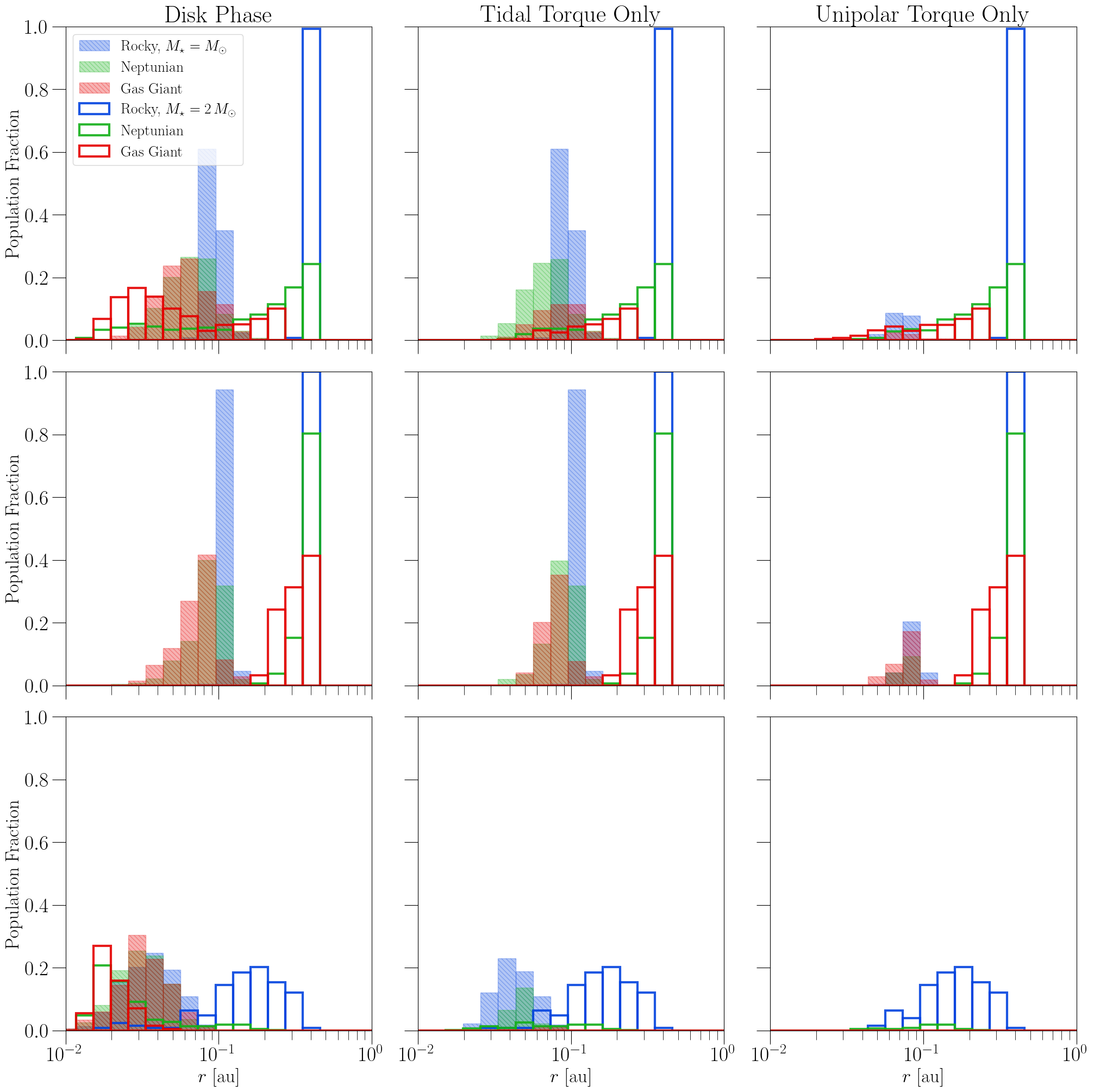}
\caption{Population fraction histograms showing the radial distributions of surviving planets categorized as Rocky (blue), Neptunian (green), and Gas Giant (red) after undergoing different migration scenarios. Results are presented for two stellar masses: 1~$M_{\odot}$ (filled/hatched bars) and 2~$M_{\odot}$ (open bars). Each row corresponds to the same disk conditions as Figure \ref{fig:planet_pop}. The simulations were run with an initial population of 10,000 planets for enhanced statistical robustness. Note that histogram values for each planetary regime do not necessarily sum to unity due to planets ending up accreted onto the host stars during migration.}
\label{fig:planet_frac}
\end{figure*}

%%% Planet population discussion
To quantitatively explore the impact of different migration mechanisms, we utilize the previously derived timescales to perform a simple population synthesis of planets orbiting around one and two solar mass stars, representative of young T Tauri and Herbig Ae/Be stars, respectively.

We first assume protoplanetary disks characterized by accretion rates of $\dot{M}=10^{-8}\,M_{\odot}/\mathrm{yr}$, persisting for a fiducial lifetime $t_{\mathrm{disk}}=10$~Myr. The temperature at 1 au is set to 317 K for T Tauri stars and 670 K for Herbig Ae/Be stars, with a radial temperature dependence of $T \propto r^{-1/2}$. Consequently, the dead-zone inner boundaries (DZIBs) are located at 0.1 au for T Tauri and 0.45 au for Herbig Ae/Be stars, where the disk temperature reaches 1000 K. The stellar dipole magnetic fields are modeled as normal distributions: $1$ kG mean at $2~R_{\odot}$ for T Tauri stars and $100$ G mean at $3~R_{\odot}$ for Herbig Ae/Be stars, each with a standard deviation of 0.3 dex. Using these temperatures and  $B_{\star}$-fields parameters, the magnetospheric truncation radius ($R_T$) (mean $B_{\star}$-values correspond to $\sim$ 13 and 6~$R_{\odot}$ for T Tauri and Herbig Ae/Be stars, respectively), disk viscosity profile ($\alpha$; see Equation \ref{eq:alphap}), and disk surface density profiles are derived self-consistently.

We then initialize a population of 1000 planets around each type of stars at the dead-zone inner boundary, with mass ratios ($q = M_p/M_{\star}$) drawn from a log-uniform distribution between $10^{-6}$ and $10^{-2}$. We assume that all planets at the beginning are located at the DZIB because it's either a preferable place for planet formation due to dust trapping \citep{Flock_2019} or a migration trap due to the corotation torque \citep{Masset_2006}.
A planet's radius can be estimated using the mass-radius relation
\begin{equation}
    \frac{R_p}{R_{\oplus}} = C \left( \frac{M_p}{M_{\oplus}} \right)^k,
    \label{eq:mass_radius_relation}
\end{equation}
where \(R_{\oplus}\) and \(M_{\oplus}\) are Earth's radius and mass being used as references, while coefficient \(C\) and exponent \(k\) are values that have been derived empirically from observational data of known exoplanets of different mass regimes \citep[see][]{Chen_2016}. These classifications are shown in \ref{tab:mass_radius_k}.

\begin{table*}[t!]
\centering
\caption{Adapted from \cite{Chen_2016}. Mass-radius relation parameters by planetary regime.}
\label{tab:mass_radius_k}
\begin{tabular}{lccc}
\hline
\textbf{Planet Regime} & \textbf{Mass Range} \((M_\oplus)\) & \textbf{Coefficient \( C \)} & \textbf{Exponent \( k \)} \\
\hline
Rocky & \( M < 2.04 \) & 1.01 & 0.28 \\
Neptunian & \( 2.04 < M < 132 \) & 0.81 & 0.59 \\
Gas Giant & \( 132 < M < 2.66 \times 10^4 \) & 17.8 & -0.04 \\
%Stellar & \( M > 2.66 \times 10^4\,M_\oplus \) & 0.88 \\
\hline
\end{tabular}
\end{table*}

During the disk phase, planet migration follows:
\begin{equation}
r_{\mathrm{end}} = R_{\mathrm{DZIB}} \exp\left(-\frac{t_{\mathrm{disk}}}{t_{\mathrm{mig}}}\right),
\end{equation}
where $t_{\mathrm{mig}}$ is the Type I/II migration timescale from Equation \ref{eq:disk-driven}. In this phase, planets ending within the truncation radius remain fixed there, since the migration timescale within the magnetosphere is longer than the disk's lifetime.

Post-disk evolution is considered by subjecting the remaining planet population separately to tidal and unipolar magnetic migration scenarios. In both cases, the star is assumed non-rotating ($\Omega_{\star}=0$), eliminating a corotation radius and leaving inward migration as the only option. This is justified since main sequence stars rotate significantly slower than their young counterparts.

After a given time \(t\), the final position \(r_{\mathrm{end}}\) of a planet, in a circular orbit, subjected torque \(\Gamma\) can be obtained by solving the differential Equation \ref{eq:rate} with the appropriate torque prescription. For tidal evolution, substituting the tidal torque (Equation \ref{eq:tidal} with $\Omega_{\star}=0$), yields the final orbital distance
\begin{equation}
r_{\mathrm{end, tidal}} = (r_p^8 - 16G M_p \tau_{\star}R_{\star}^5t)^{1/8},
\end{equation}
where $t=1$ Gyr for one solar mass stars and $100$ Myr for two solar mass stars, accounting for the more massive star's shorter lifetime.
%\todo{You may want to at least give the equations on how to derive this. } - DONE: extra sentence about differential equation

Similarly, for migration driven by the unipolar magnetic torque (Equation \ref{eq:unipol_torque}), leads to
\begin{equation}
r_{\mathrm{end, U}} = (r_p^6 - 96R_p^2B_{\star}^2R_{\star}^6M_p^{-1}\mathcal{R}_{\star}^{-1}t)^{1/6},
\end{equation}
where we assume that the stellar field \(B_{\star}\) has decayed and is a hundred times weaker than during the disk phase (averaging 10 and 1 Gauss, respectively). Planets reaching stellar radii are considered accreted and thus removed from the population.
%\todo{This equation is not correct, it does not have t. Please double check equations 38 and 39.}

The top row of Figure \ref{fig:planet_pop} illustrates these outcomes for both stellar types under the fiducial conditions.
During the disk phase, gas giants in the range of Saturn to Jupiter mass ($10^{-4} \lesssim q \lesssim 10^{-3}$) experience significant inward migration, whereas more massive giants (which are in the Type II regime of disk-driven migration) and the less massive Neptunian and rocky planets ($q \lesssim 10^{-4}$) migrate at more modest rates.
The exact radial pile-up is set by the magnetospheric truncation radius, $R_T\!\propto\!B_\star^{4/7}$: stars in the upper tail of our assumed 0.3 dex spread in dipole strength halt planets farther out, while weaker-field stars allow the same planets to spiral further in before reaching the magnetosphere's boundary.

%Because the subsequent tidal ($t_{\rm tidal}\!\propto\!r_p^8$) timescale is a extremely steep functions of distance, this dispersion in $R_T$ maps directly onto planetary fates after the disk disperses: planets stranded at the larger $R_T$ values of high-$B_\star$ systems survive being engulfed by the tidal torque, with lower mass planets being more favored for survival than those of higher mass ($t_{\rm tidal}\!\propto\!q^{-1}$).
%Similarly, the unipolar magnetic ($t_{\mathrm{mag, U}}\!\propto\!r_p^6$) timescale is also a quite steep function of distance, with a difference being the presence of the strong inverse dependence on the remaining stellar B-field  ($t_{\mathrm{mag, U}}\!\propto\!B_{\star}^{-2}$). As a consequence, too high of a remnant stellar field ends up removing planets, despite them being initially stranded at large $R_T$ values. This end result being the disappearance of close-in planets of medium mass.

Because the tidal decay timescale scales steeply with orbital distance ($t_{\rm tidal}\!\propto\!r_p^{8}$) and inversely with planet mass ($t_{\rm tidal}\!\propto\!q^{-1}$), this dispersion in $R_T$ maps directly onto post-disk survival. Planets stranded at larger $R_T$ in high-$B_\star$ systems are largely immune to tidal engulfment, especially at lower masses, whereas those parked closer in around weak-field stars are readily removed.  
Unipolar magnetic migration behaves similarly steeply in $r_p$ ($t_{\mathrm{mag,U}}\!\propto\!r_p^{6}$) but carries an extra inverse dependence on the residual stellar field ($t_{\mathrm{mag,U}}\!\propto\!B_\star^{-2}$). Consequently, planets around stars that retain strong remnant fields can still be cleared out, even if they started from comparatively large $R_T$, leading to a gradual, preferential loss of gas giants.
% a chasm is being opened gradually

Note that, to demonstrate the tidal interaction more clearly, we have kept the strong tidal interaction as in young phases ($\tau_{\star}=1$ s, or $Q\sim10^5$ for planets with 7 days periods). The realistic tidal interaction could be significantly weaker \citep{Lee_2017}, restricting tidal consumption to even smaller orbits. Meanwhile, the unipolar interaction requires forming a closed circuit between the planet and the star, which may not always be the case (see discussions in \citealt{Laine_2012}). Thus, we consider our cases the limiting cases with the strongest possible tidal and magnetic effects.

The middle row of Figure \ref{fig:planet_pop} demonstrates outcomes for a shortened disk lifetime of $1$ Myr. Because planets have little time to migrate before dispersal, subsequent tidal and magnetic migration is correspondingly weakened. The bottom row depicts a scenario with a higher accretion rate ($10^{-7} M_{\odot}$/yr), where higher disk surface densities lead to more significant migration, populating the innermost region and thus setting up the planets for rapid post-disk removal.
% sub-neptunes and rocky planets favored in last case
%\todo{Mention the effect of the stellar B-field on a planet's fate.} - DONE

Figure \ref{fig:planet_frac} recasts the same population synthesis result as fractional radial histograms, separating the surviving planets into our three mass regimes (Table \ref{tab:mass_radius_k}).
In the fiducial disks (top row), while rocky planets linger at the dead-zone edge, gas giants and Neptunians of both stellar masses have already drifted inward to the truncation radius by the time the gas disperses. Their dispersion in $r$ is determined by the dispersion of $R_T$ and thus $B_{\star}$ of the stellar population. Tides then remove almost every giant and nibble only at the innermost Neptunians. The unipolar torque is much harsher: it wipes out everything except low-mass rocky worlds and a handful of super-giants in the solar case and is actively eroding giants and Neptunians in the higher-mass host, yet still spares its rocky population.

Next, with a 1 Myr disk phase (middle row), very little migration occurs before dispersal, so the initial peaks are narrow and distant; tides trim only the closest giants, but a strong unipolar circuit again clears nearly all giants and Neptunians around the solar-mass host. The two-solar-mass systems, parked farther out and endowed with a weaker remnant field, remain largely unchanged.

The high-accretion run (bottom row) drives every planet headlong into the cavity, producing tight stacks at the truncation radius. Tides promptly consume all giants and most Neptunians for both stellar masses. Unipolar torques delete the whole planet population around the solar-mass star and leave a remnant population of rocky planets around the heavier star.
%\todo{You need to describe the figure and discuss the figure. Guide the readers to understand the figure and make points. Ideally you need to go over each panel, or at least each row.} - DONE

Taken together, these variations highlight the sensitivity of the final planet population to conditions during the disk phase. As for the post-disk phase, their outcomes are governed by poorly constrained parameters. The tidal interaction is quite uncertain, reflected by the tidal quality factor $Q$ (which sets the tidal time lag $\tau_{\star}$). For solar-type stars, values of $Q \sim 10^{5-8}$ are commonly assumed \citep[e.g.,][]{Lee_2017,wei2024magneticfieldgasgiant}, but the exact value can heavily affect orbital migration rates and the resulting planet demographics; lower Q-values (higher tidal dissipation) accelerate inward migration, rapidly removing close-in planets, while higher values allow more planets to survive at short orbital periods. This value can fluctuate during a star's lifetime by several orders of magnitude, being strongest (lower Q-value) during the star's early history  \citep[see e.g.,][]{Gallet_2017}.

Similarly, a star's surface \(B_{\star}\) is the strongest early on. Whether unipolar induction can operate depends on if the planet loses its own magnetosphere and the planet's resistivity is small enough. Since the dipolar regime is more expected \citep[see e.g.,][]{Strugarek_2017}, we have shown the most unfavorable case for planet survival in Figures \ref{fig:planet_pop} and \ref{fig:planet_frac}.

In the past, tidal interactions have been shown to widen orbital spacings for close-in planets due to preferential inward migration and mergers at smaller radii \citep[see][]{Lee_2017}. Moreover, photoevaporation and magnetic drag may also significantly shape the population of rocky planets at short periods, potentially explaining observed boundaries and deficits \citep[see e.g.,][]{Lee_2025}. Our simplified analysis aligns broadly with these findings, highlighting the combined influence of disk truncation, tidal, and magnetic torques in shaping the distribution of close-in planets. 

On the other hand, our analysis emphasizes that planetary positions at the end of the brief disk phase largely affect their demographics over billions of years, especially for low-mass planets which are more favored to retain their positions. Thus, exoplanet demographics studies could constrain planet formation and migration processes. Our results are broadly consistent with two recent exoplanet demographic works by \cite{Mendigutia_2024} and \cite{Sun_2025}.

\cite{Mendigutia_2024} suggest that hot Jupiters are halted by the magnetosphere instead of the dust sublimation front in disks. This is based on their findings that hot Jupiters around intermediate-mass stars (1.5-3 $M_{\odot}$) tend to be closer to their hosts than the inner dust disk, and there is no correlation between orbit sizes and stellar luminosities expected if the dust sublimation front is responsible. Their result is consistent with our fiducial model that Jupiter mass planets can migrate within the inner MHD turbulent disk and park at the magnetospheric truncation radius. On the other hand, we caution that our model did not consider high eccentricity migration (e.g. planet-planet scattering), which may play important roles in hot-Jupiter migration.

Meanwhile, \cite{Sun_2025} indicate that super-Earths and sub-Neptunes are predominantly curtailed by the dust sublimation radius (see also \citealt{Flock_2019}), since these planets are generally farther away from the central star when the star is hotter. This is also consistent with our fiducial model where low mass planets cannot migrate into the inner MRI active disk and stay at the deadzone inner boundary. The DZIB is determined by the thermal ionization, which has a temperature around $\sim$1000 K \citep{Desch2015}. This temperature is close to the dust sublimation temperature $\sim$ 1000-1500 K, and thus the dead zone inner boundary coincides with the dust sublimation radius.

\subsection{Planet Pairs' Stability}

\begin{figure}[t!]
\centering
\includegraphics[trim=0mm 0mm 0mm 0mm, clip, width=\columnwidth]{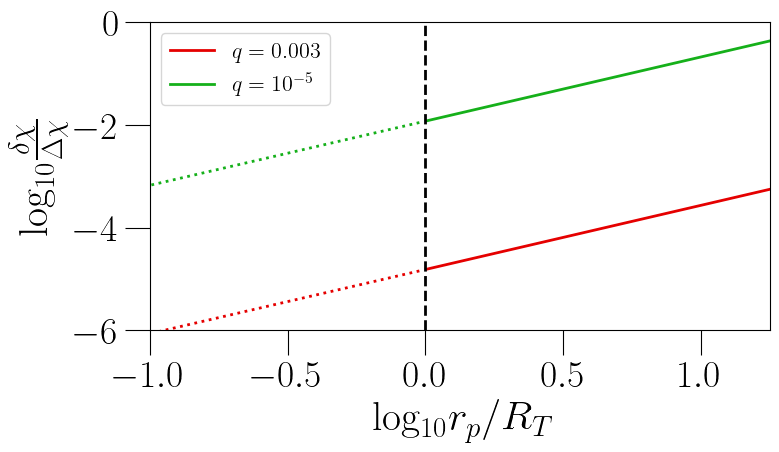}
\caption{Turbulent disruption of mean-motion resonance for close-in planet pairs at various orbital distances. A resonance remains stable under stochastic forcing if $\delta \chi / \Delta \chi \ll 1$. Inside the magnetically truncated cavity, disk density is significantly reduced.}\label{fig:resonance}
\end{figure}

During the disk phase, once embedded planets reach the magnetospheric truncation radius $R_T$ they stall. Because planets of different masses and starting locations converge on $R_T$ at slightly different times, their final spacings are small and mean-motion resonances are readily established \citep[see e.g.,][]{Goldreich_1965,Allan_1969,Allan_1970,Sinclair_1970,Sinclair_1972}. These can be broken by turbulent fluctuations if the diffusive forces acting on planetary orbits become large enough. Although turbulence operates over long timescales, these random perturbations may still drive planet pairs out of resonance \citep[see e.g.,][]{Adams_2008,Leocanet_2009,Ketchum_2011}. \cite{Batygin_2017} gives a criterion for resonance disruption of the form
\begin{eqnarray}
    \frac{\delta \chi}{\Delta \chi} \sim \frac{h}{20} \frac{M_{\star}}{M_{p,1} + M_{p,2}} \sqrt{\frac{3\alpha}{f_{\mathrm{res}}}} \nonumber \\
    \times \left[ \frac{\Sigma \langle r_p \rangle^2}{k_{\mathrm{res}}M_{\star}} \sqrt{\frac{\Sigma \langle r_p \rangle^2}{M_{p,1} + M_{p,2}}} \right]^{1/3} \gtrsim 1,
\end{eqnarray}
where $m_1$ and $m_2$ are the planetary masses, $f_{\mathrm{res}} = 1$ and $k_{\mathrm{res}} = 3$ are resonance constants, and $\langle a \rangle$ is the average semi-major axis. In this expression, $\delta \chi$ describes the broadened resonant width due to turbulence, while $\Delta \chi$ is the resonance stability threshold.

In this parking-zone environment, the survival of resonances is governed chiefly by the planet-star mass ratio, with a weaker dependence on (local) disk surface density. Systems whose total mass ratios fall below $q = 10^{-5}$ are therefore the most vulnerable to stochastic break-up. In Figure \ref{fig:resonance} we apply the \citet{Batygin_2017} criterion and show that the ratio
$\delta\chi/\Delta\chi$ remains well below unity ($\ll 1$) for
a $q = 0.003$ pair.  Even for a $q = 10^{-5}$ pair,  $\delta\chi/\Delta\chi$ stays below unity.

This theoretical expectation of resonant planet pairs is supported by the planet demographics study from \citet{Dai_2024}. Surveying every age-dated multiplanet system, they find that young systems ($<$100 Myr) retain first-order resonances in 70\% of adjacent pairs, whereas only 15\% of mature systems ($>$1 Gyr) do so. They argue that resonant chains are forged during disk-driven migration and dissolve on $\sim100$ Myr timescales after gas dispersal, when dynamical instabilities, stochastic forcing, and relatively weak post-disk torques overpower tidal or magnetic damping. 
This observation is consistent with our results that MRI turbulence cannot break the resonances so that resonant planets could be abundant.  We note that some planetary systems in \citet{Dai_2024} may reside inside the deadzone (as discussed above), so that their resonance stability requires future theoretical work studying the deadzone turbulent structure.

\subsection{Model's Caveats}

Although our 3-D simulations incorporate the magnetorotational instability (MRI) and a magnetically truncated inner disk, several limitations remain. First, we exclude stellar rotation, despite its capability to shift the corotation radius substantially and reshape the inner disk \citep[see][]{Zhu_2025}.
Future simulations should explore a range of initial stellar rotation periods and study planet migration in such disks. Once stellar rotation is incorporated, the magnetized stellar wind must also be taken into account, as it removes angular momentum from the star and affects disk-star coupling \citep[e.g.,][]{Matt_2005,Cranmer_2008}. In our present configuration, because the stellar temperature and density are not fully realistic, the density in the stellar wind region would often reach the imposed floor value, artificially minimizing any torque contribution from that region.

Second, we fix the star’s magnetic dipole along the simulation’s z-axis (perpendicular to the disk's midplane), as would be the canonical aligned-dipole case if the star were spinning, and neglect potential large-scale external fields. We caution, however, that real T Tauri stars often exhibit oblique or multipolar fields, and that large-scale background fields can thread the system \citep[see e.g.,][]{Donati_2007,Lankhaar_2022}. General relativistic MHD simulations indicate that misaligned or additional background fields can introduce nontrivial effects in disk dynamics, jet formation, and stellar angular momentum loss \citep[e.g.,][]{Romanova_2012,Romanova_2015,Parfrey_Tchekhovskoy_2023,Murguia-Berthier_2024}.

%Second, we assume the star’s magnetic dipole aligns with its rotation axis and neglect potential large-scale external fields. General relativistic MHD simulations indicate that misaligned fields or additional background fields can introduce nontrivial effects in disk dynamics, jet formation, and stellar angular momentum loss \citep[e.g.,][]{Romanova_2012,Romanova_2015,Parfrey_Tchekhovskoy_2023,Murguia-Berthier_2024}. %\todo{Some Romanova's papers also mentioned its effect.} - DONE

Third, we do not include full radiative transfer or realistic thermodynamics. In reality, the temperature structure in the magnetosphere is shaped by X-ray heating, magnetic reconnection, and wave dissipation, among other processes \citep[e.g.,][]{Hartmann_2016}. Capturing all these processes in a global MHD setting remains a challenge.

Fourth, our simulations treat planets as gravitational potentials, with a smoothing length, but without a solid surface, which allows gas parcels to reach artificially high velocities very close to the planetary center. This can lead to numerical instabilities or unphysical flow speeds. Future work should employ more sophisticated planetary boundary conditions or a sub-grid model to prevent these issues \citep[e.g.,][]{Kley_1999}.

Fifth, with the limited orbital time (tens of orbits) in our 3-D simulations, we cannot measure the mean migration torque (e.g. Type I/II) among the highly fluctuating turbulent torques. Although previous shearing-box unstratified MHD simulations show that Lindblad resonances could launch spirals in turbulent disks \citep{zHU_2013}, the disk structure around the magnetic truncation radius is much more complex due to its much higher magnetization and other instabilities (including interchange instability). Recent 2-D simulations with driven turbulence have also demonstrated that Type I migration can be reduced in highly turbulent disks \citep{Wu_2024}. Thus, long-time scale 3-D MHD simulations are needed in future to study if the traditional Type I/II migration rate can still be applied over the lifetime of the turbulent disk. 

Finally, we set a disk aspect ratio of $h(r=R_0)=0.1$ in our 3-D simulations, which is thicker than the $\lesssim 0.05$ commonly inferred for the inner protoplanetary disks \citep[e.g.,][]{Crida_2006}. Maintaining adequate resolution in a thinner disk would increase computational cost by at least an order of magnitude. However, thinner disks also allow massive planets to clear gaps more effectively, which can alter migration timescales and overall disk structure. On the other hand, we do consider a more realistic disk aspect ratio in our analytical estimate in the Discussion Section (\S \ref{sec:disc}).

%\todo{ACS: Uncertainties in tidal and magnetic torques but bottom line, the disk-phase sets the end result}

\section{Conclusions}\label{sec:conclusion}

In this work, we performed high-resolution 3-D ideal MHD simulations and semi-analytical modeling to investigate how young planets migrate in the innermost regions of protoplanetary disks. Our simulations considered planet masses ranging from super-Jovian (\(q=0.01\)) to super-Earth masses (\(q=10^{-4}\)) embedded within turbulent disks governed by the magnetorotational instability (MRI).

From direct analysis of our simulated torque time series, we observed that stochastic torques driven by MRI turbulence exhibit shorter correlation times (\(\tau_c \sim 0.1\) local orbital periods) than previously estimated, implying rapid decorrelation of turbulent structures, and indicating that the turbulent fluctuations felt by the planet lose “memory” quickly. At the same time, we observed consistent periodic signals that recur on each orbital revolution, suggesting that large, long-lived structures in the outer disk modulate the planet’s torque.

To reproduce this combination of short-lived stochastic torque and low-level periodicity, we introduced a synthetic model (Equation \ref{eq:markov_sin}) combining a first-order Markov process with a small-amplitude sinusoidal term. We further quantified the stochastic torque with the local disk turbulence strength (Equation \ref{eq:cd-rad}).

Using our simulated disk structure, we discuss various planet migration mechanisms at the inner MRI active disk.  Although MRI turbulence exerts strong, fluctuating torques, the net diffusional change on a planet's orbital radius is minimal. Type I/II migration dominates in the disk, although it is still less effective with such a low disk surface density. Only giant planets, which cannot open gaps in such strongly turbulent disks and are in the Type I regime, can traverse the inner MRI turbulent disk within typical disk lifetimes.  Meanwhile, low-mass planets may migrate inwards in the deadzone but fail to migrate in the inner MRI active zone and stall near the deadzone inner boundary (DZIB).

Turbulence in the low-density disk is unable to break the resonant planets during the disk's lifetime, and thus young planets in resonances may be abundant.

Once a giant planet arrives near the magnetospheric truncation radius, they become primarily influenced by aerodynamic drag, tidal, and magnetic torques. Whether these planets remain trapped at the truncation radius or migrate rapidly inward depends on the strength of these effects and
the relative positions of the truncation radius ($R_T$) and the stellar corotation radius $R_C$.
With sufficient tidal and magnetic torques, if $R_C > R_T$, massive planets may spiral inward, whereas if $R_C < R_T$, they tend to remain trapped.

We further modeled a simple population of planets evolving under these mechanisms around both T Tauri and Herbig Ae/Be stars. During the disk phase, planets that are either too low in mass or too distant migrate slowly and remain near their formation radius. Only intermediate to massive planets experience significant inward migration, potentially reaching the magnetospheric truncation radius before the disk disperses.

Post-disk evolution is dominated by stellar tidal and magnetic torques, but these torques mainly affect planets that are already close-in. These preferentially remove hot gas giants over Gyr timescales (even faster on more massive stars), while rocky and Neptunian planets further out remain largely unaffected. On the other hand, we note that these processes are poorly understood and we likely overestimate their effects with our chosen tidal and magnetic parameters. Overall, planetary positions at the end of the brief disk phase largely
affect their demographics over billions of years. Thus, exoplanet demographics
studies could constrain planet formation and migration processes.

Our model that giant planets can migrate to the magnetospheric truncation radius while low mass planets stall at the dead zone inner boundary is consistent with 
recent planet demographic works by \cite{Mendigutia_2024} and \cite{Sun_2025}. Furthermore, resonant planets that are in MRI active inner disks could be abundant, which may have observational implications, as studied in \cite{Dai_2024}.

\begin{acknowledgements}
    Simulations are carried out using NASA Pleiades supercomputer. A. C. S. and Z. Z. acknowledge support from NASA award 80NSSC22K1413 and NSF award 2429732 and 2408207. The authors thank Vassili Desages for discussion and contribution at the early stage of the project. 
\end{acknowledgements}

\appendix

\section{Disk Sectors Torque Analysis}\label{sec:appendix}

\begin{figure*}[t!]
\centering
\includegraphics[trim=0mm 0mm 0mm 0mm, clip, width=7.in]{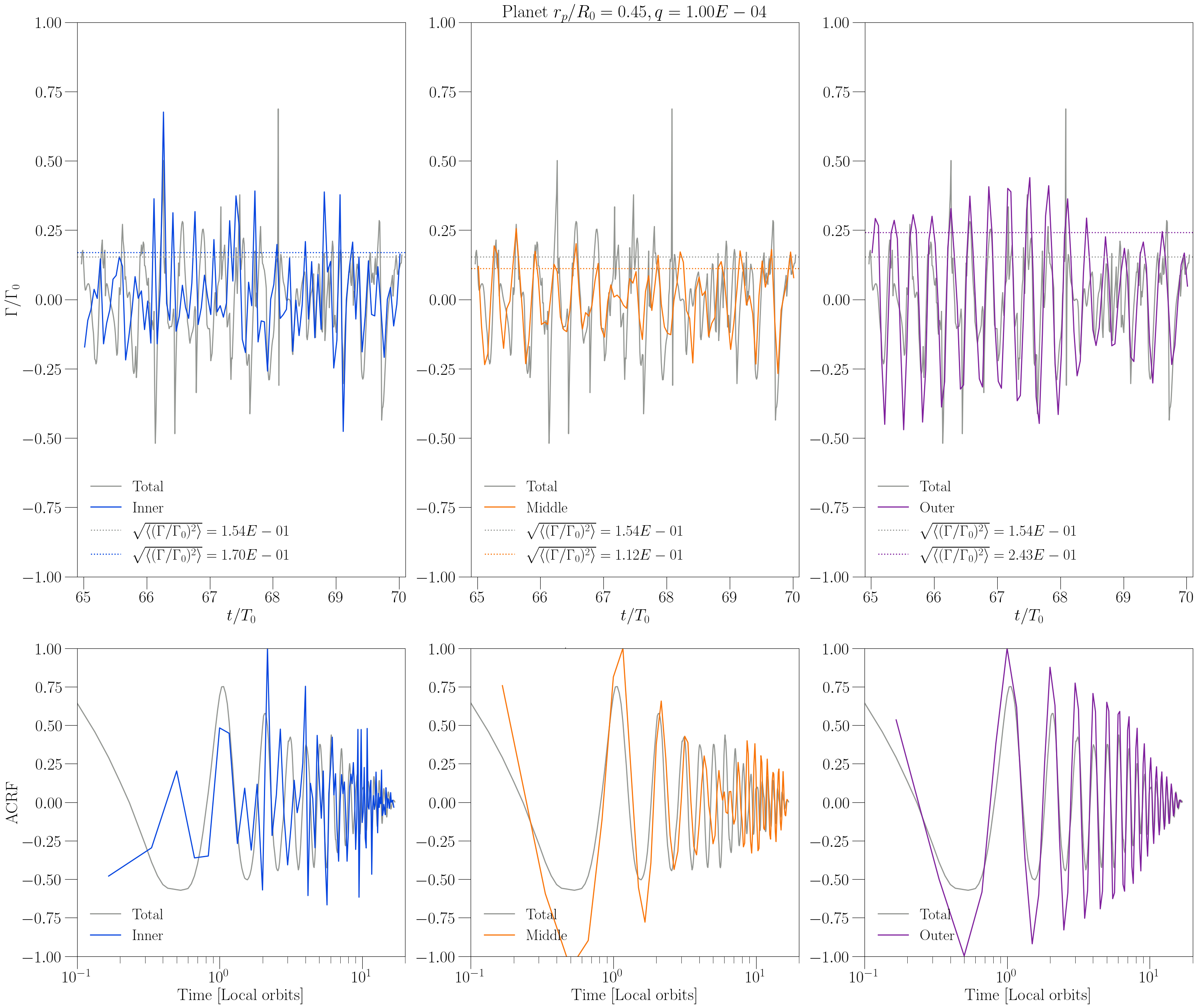}
\caption{\textbf{Top:} Normalized planetary torque felt by the second planet in case B (\(r_p = 0.45\), \(q=10^{-4}\)), arising from different annuli of the accretion disk versus the total. \textit{Inner}, \textit{Middle}, and \textit{Outer} sectors correspond to annuli at \(r \leq 0.8\), \(0.8 < r \leq 1.5\), and \( r > 1.5\), respectively. \textbf{Bottom:} Autocorrelation functions of the sectoral torque.}\label{fig:segment-0}
\end{figure*}

\begin{figure*}[t!]
\centering
\includegraphics[trim=0mm 0mm 0mm 0mm, clip, width=7.in]{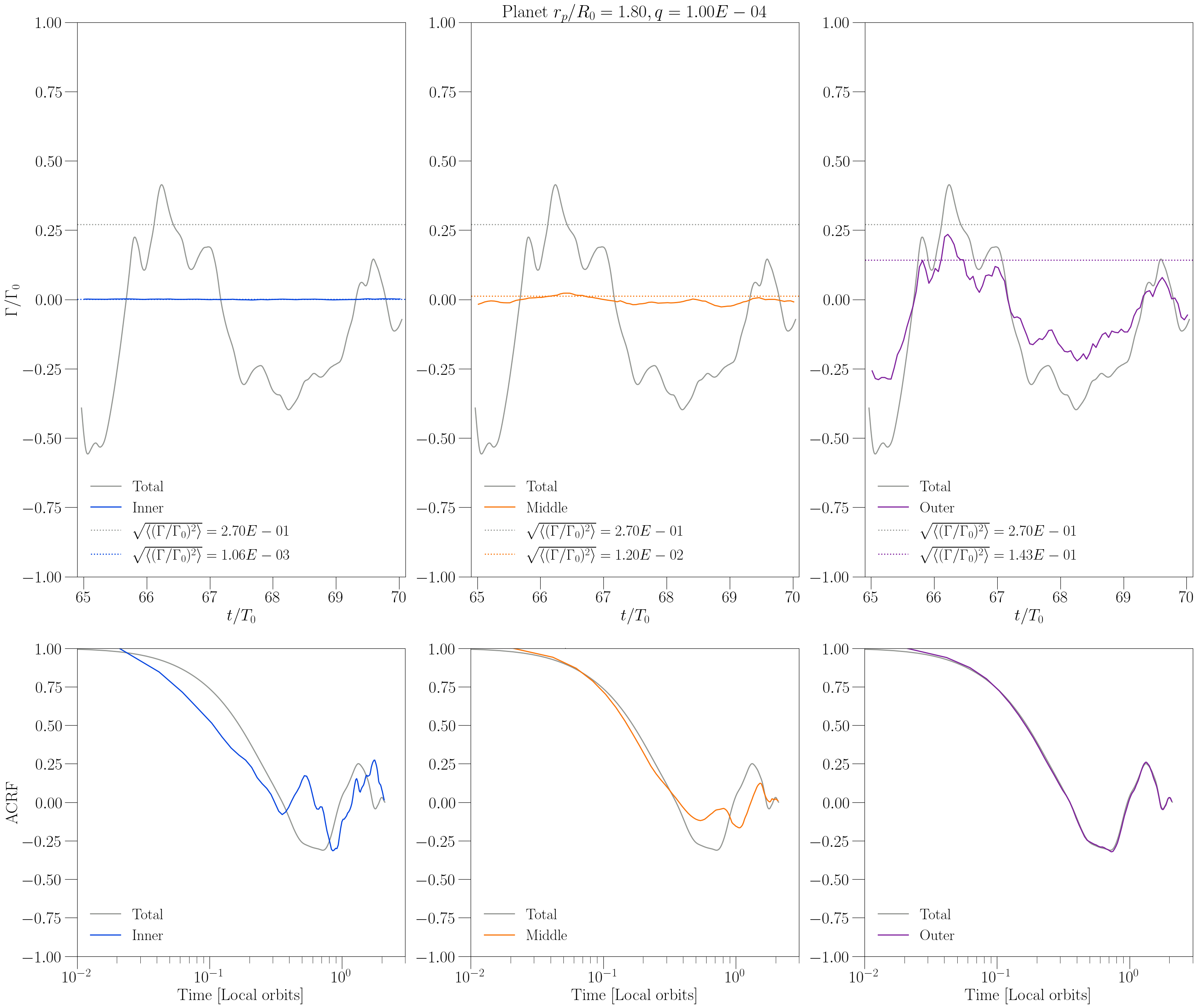}
\caption{\textbf{Top:} Normalized planetary torque felt by the second planet in case B (\(r_p = 1.8\), \(q=10^{-4}\)), arising from different annuli of the accretion disk versus the total. \textbf{Bottom:} Autocorrelation functions of the sectoral torque.}\label{fig:segment-2}
\end{figure*}

Here we provide further examples related to Figure \ref{fig:segment-1} by splitting the disk into three radial sectors: $r \leq 0.8$, $0.8 < r \leq 1.5$, and $r > 1.5$. Figures \ref{fig:segment-0} and \ref{fig:segment-2} correspond to planets at $r_p = 0.45$ and $r_p = 1.8$ (both with $q=10^{-4}$) in case B. The torque time series and the corresponding autocorrelation functions (ACRFs) confirm that the periodic structure primarily originates from the outer disk. This pattern is most evident for the innermost planet, which shows pronounced peaks in its outer-disk torque ACRF. For the planet near $r_p=1.8$, fewer orbits are completed, but the outer disk’s ACRF also matches the overall torque behavior shown in Figure \ref{fig:autocorrelation}.

%%% Segmented disk torque planet 0

%%% Segmented disk torque planet 2

\bibliography{sample631}{}
\bibliographystyle{aasjournal}

\end{document}